%% file: main.tex
\documentclass[11pt]{article}
\usepackage{natbib}
\usepackage[bottom=1in]{geometry}
\usepackage[colorlinks=true, linkcolor=black, urlcolor=blue, citecolor=blue, anchorcolor=blue]{hyperref}
\usepackage{url}
\usepackage{amsmath,amssymb,amsthm, amsfonts}
\usepackage{tabularx}
\usepackage{multirow}
\usepackage{colortbl}
\usepackage{diagbox}
\usepackage{pdflscape}
\usepackage{rotating}
\usepackage{tikz}
\usetikzlibrary{positioning,arrows.meta,fit,backgrounds}
\usepackage{pgfplots}
\usepgfplotslibrary{groupplots}
\usepackage{mathrsfs}
\usepackage{xifthen}
\usepackage{amsmath}
\usepackage{xparse}
\include{commands}

\usepackage[font=small,labelfont=small]{caption}
\usepackage[subrefformat=parens, labelfont=up]{subcaption}

\date{}
\begin{document}
\title{Policy relevance of causal quantities in networks}
\maketitle

\begin{center}

  \thispagestyle{empty}
  
  \begin{tabular}{cc}
    Sahil Loomba\textsuperscript{\dag} and Dean Eckles\textsuperscript{\dag,\ddag,*}
   \\[0.25ex]
   {\small \textsuperscript{\dag} MIT Institute for Data, Systems, and Society} \\
   {\small \textsuperscript{\ddag} MIT Sloan School of Management} \\
  \end{tabular}

  \vspace*{0.1in}
    {\footnotesize\textsuperscript{*}eckles@mit.edu}
    \vspace*{0.1in}
    
    {\footnotesize 9 March 2026}
  \vspace*{0.4in}
  
\begin{abstract}
In settings where units' outcomes are affected by others' treatments, there has been a proliferation of ways to quantify effects of treatments on outcomes, including via indirect exposure to other units' treatments. Here we consider two properties we might want estimands to have: being interpretable as summaries of unit-level effects, and being relevant to choice of a policy governing treatment assignment. We characterize many estimands as involving one of two orders of averaging over units in a population and over treatment assignments under a policy. The more common representation often results in quantities that are insufficient for optimal policy choice. This occurs because these quantities summarize outcomes under homogeneous exposure to treatment, but even homogeneous policies often lead to heterogeneous exposures. The other representation often yields quantities that lack an interpretation as summaries of unit-level effects. We argue that, among various estimands, the expected average outcome, which averages over units and treatment assignments in either order, deserves further attention from researchers. This estimand, or contrasts among these estimands under different policies, is both a summary of unit-level effects and is sufficient for optimal policy choice with utilitarian welfare.
\end{abstract}
\vspace*{0.15in}
\end{center}

\hspace{10pt}
  \small	
  \textbf{\textit{Keywords: }} {causal inference, interference, social networks, social spillovers, counterfactual policy evaluation}

\section*{How should we quantify the effects of an intervention?}
In the case of evaluating the effects of a binary treatment, such as through a randomized experiment, the average treatment effect (ATE) is often the primary quantity of interest (or \emph{estimand}). It is sometimes sufficient for decision making.
Consider a setting where the outcome of each unit $i\in[n]$ depends on its own treatment $Z_i$, so that we observe $Y_i \triangleq y_i(Z_i)$; that is, there is ``no interference'' \citep{cox58}.\footnote{We use upper case letters to denote (observed) random variables and boldface letters to denote vectors.} Say that a decision maker aims to maximize utilitarian welfare, $W(\vect{z}) \triangleq \avgpi{i\in [n]}{y_i(\vect{Z})}$, where throughout we use $\avgpi{i\in [n]}{h_i}$ to denote averages over a population of units.
Then, when deciding whether to apply a costless treatment to an entire population or not, the decision maker can simply look at the sign of the $\text{ATE} \triangleq \avgpi{i\in [n]}{y_i(1) - y_i(0)}$. 
Even with a costly treatment, the ATE may remain a sufficient summary of the potential outcomes for decision-making, such as choosing among policies governing randomized allocation of a treatment. Given known costs of treatment (which may be nonlinear in how many units are treated), the ATE tells the decision maker what proportion of the population to treat at random in order to maximize expected welfare.\footnote{Of course, we typically do not know the ATE, but rather have sample data, such as that arising from a randomized experiment or observational data.
Then we have to decide how to make choices under uncertainty and/or ambiguity; see, e.g., \citet{manski2004statistical}. In this article, we focus solely on what quantities are of interest, neglecting estimation and inference.}
More generally, with an ordinal or continuous treatment, averages of potential outcomes under each possible ``dose'' constitute a dose--response curve and can allow selecting the optimal dose or distribution of doses.

By contrast, in settings where the units' outcomes are affected by others' treatments, how to quantify effects has been less clear. 
This includes how to study important questions about how changes to communication technologies affect the spread of information, how environmental policy affects public health, and how recommendation and matching algorithms affect market outcomes.

Here we focus on the case of interventions in a network, which, while most obviously applicable to social networks, can also be used to represent interference in other settings (e.g., bipartite networks representing which regions are potentially exposed to pollution from which power plants).
In a social network where the outcome for one unit might depend on the treatment of all other units, there are many possible comparisons to make: with a binary treatment, the table of all potential outcomes is $n \times 2^n$ (Figure \ref{fig:overview}, top left), and there are many ways to summarize and contrast these entries. Naturally, there has been a proliferation of proposed estimands. Many of these are intended to capture various notions of not only direct effects (effects of a unit's own treatment on itself), but also spillover effects. These spillover effects include various versions of how units are affected by other units' treatments or, to put it the other way around, how some units' treatments might affect other units.

To guide what quantities we should develop estimators for and what empirical researchers should report, it may be helpful to consider two properties we would like estimands to have --- properties that the ATE and dose--response curves satisfy in the absence of interference. First, we often want estimands to be summaries of unit-level causal effects. To spell out and generalize this idea, we say an estimand is \emph{unit-level causal} with respect to some policy space if it can be represented as a convex combination of unit-level contrasts of each unit's outcome under different, possibly unit-specific, policies in that space. This thus includes quantities like the ATE, but also effects of stochastic interventions \citep{munoz2012population,kennedy2019nonparametric,chin2022evaluating}. For example, we may want a quantity that is an average across units of the effect of some increase in exposure for each unit. 
Second, we often want estimands that can inform how to assign treatments. More specifically, we may want estimands that are \emph{sufficient for policy choice} in that sense that they can be used to choose a policy, from some policy class, that assigns treatments to units so as to maximize expected welfare.\footnote{Throughout the main text, we focus on utilitarian welfare with possibly costly treatments.}

\begin{figure}
    \centering
    \includegraphics[width=0.75\linewidth]{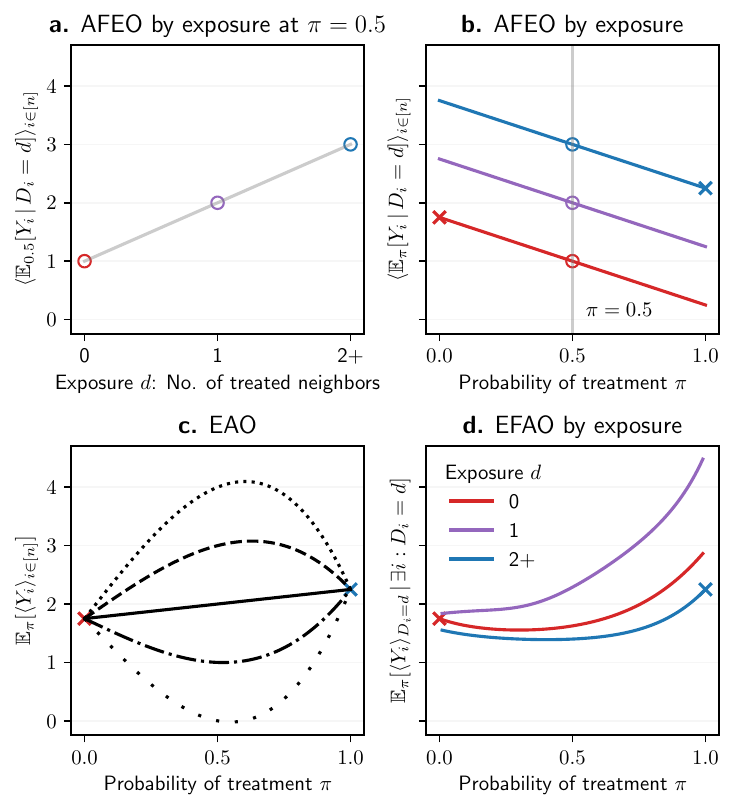}
    \caption{Different kinds of dose--response curves.
    \emph{(a)} Average focal expected outcomes (AFEOs) for each unique exposure evaluated when units are treated with probability $\pi = 0.5$. Here, $d$ is the number of treated neighbors of each unit. 
    \emph{(b)} The same AFEOs as in panel (a) but as a function of the policy $\pi$. The circles correspond to the AFEOs shown in panel (a). That these AFEOs are not constant in $\pi$ is a consequence of the exposure map being misspecified.
    \emph{(c)} The expected average outcome (EAO) as a function of the homogeneous Bernoulli policy; several such curves are all compatible with the AFEOs in panels (a, b).
    \emph{(d)} Expected focal average outcomes (EFAOs) as a function of the policy $\pi$ corresponding to the solid EAO curve in the center. Comparisons of the EFAOs for a given $\pi$ characterize the outcomes of units with different numbers of treated neighbors under that policy, but these comparisons are not summaries of unit-level causal effects.
    \emph{Notes:} Looking at the AFEO curve in panel (a) one might be tempted to conclude that the average outcome increases as more units are treated, but that question is answered by the EAO curves in (c). However, quite different EAO curves are compatible with the same AFEOs, some of which are not even monotonic --- as shown by the solid, dashed, dotted, dash-dotted, long-dotted EAO curves. In other words, the EAO is generally not identified by the AFEO. Furthermore, looking at the AFEOs in panel (a), one might be tempted to conclude that, under a given design $\pi$, the expected average outcome of exposure $2+$ units is more than that of exposure $1$ units, but that question is answered by the curves in panel (d). In fact, there exists no i.i.d. Bernoulli design for which the expected average outcome of exposure $2+$ units is larger than that of exposure $1$ units.}
    \label{fig:aoe_vs_eao}
\end{figure}

In the simple ``no interference'' case considered before, there is a single estimand (ATE; or, with a many-valued treatment, a set of estimands) that is both (a) a summary of unit-level causal effects and (b) sufficient for optimally choosing a policy that assigns treatment homogeneously. With interference, many appealing estimands that have one property will lack the other. Figure \ref{fig:aoe_vs_eao} shows different kinds of dose--response curves available here. One (Figure \ref{fig:aoe_vs_eao}a) characterizes units' average outcomes as function of their local exposure to treatment. Another (Figure \ref{fig:aoe_vs_eao}c) instead gives average outcomes as a function of a parameter indexing different policies (e.g., Bernoulli probability of treatment). And variations on this latter kind of curve (Figure \ref{fig:aoe_vs_eao}d) characterize how the average outcomes of units with different relationships to treated units vary with the policy parameter.
In what follows, we elaborate on the properties and relationship between these two types of estimands.
These estimands can be seen as involving one of two orders of averaging over units and possible treatment assignments. Here we will describe these two ways of averaging, noting that one often results in quantities that are insufficient for optimal choice of policies. The other often results in quantities that are not unit-level causal, but we argue that additional attention is due, as even then they may be applicable for policy choice. In particular, both ways of averaging generally coincide for one estimand --- namely the expected average outcome --- that we argue is both policy-relevant and causally interpretable.

\section*{Averaging over assignments, then over units}
When statisticians and empirical researchers recognize that interference may be present, it is common to try to characterize how units respond not only to their own treatment but also to neighbors' treatments. This idea has been formalized by introducing an exposure map \citep{aronow2017interference,manski2013identification}. An \emph{exposure map} $d$ maps all possible treatment assignments $\{0, 1\}^n$ into \emph{exposures} for each unit. For example, we might let the exposures for unit $i$ be all possible combinations of whether $i$ is treated and the number of $i$ neighbors who are treated. Let a \emph{focal map} be the special case of a binary exposure map that indicates which units have a particular exposure (e.g., untreated units with exactly one treated neighbor). (We leave formal definitions and results to the Appendix, but reference specific results in the main text.)

Given such an exposure map, it is natural to summarize outcomes under these different exposures. Among such summaries, most common are estimands that can be represented as first averaging over treatment assignments within units and then over units.
Considering the case of something we might label a spillover effect, we can represent the first averaging step as marginalizing over $i$'s potential outcomes according to a distribution over the treatment of all units conditional on $i$ having different number of treated neighbors (Figure~\ref{fig:overview}, top right).\footnote{
    In some cases, it is straightforward to instead define a variation on this where this distribution is not conditional on the exposure. For example, if exposures only consist of a unit's own treatment (i.e. $d(\vect{Z}) = z_i$), then we can instead marginalize over the unconditional distribution of $\vect{Z}_{-i}$; this defines what \citet{savje2021ateunknown} call an expected ATE. If treatment assignment is independent across units, then this coincides with the AFEOs discussed here. Related quantities appear in many other papers \citep[e.g.,][]{hudgens2008toward,vanderweele2011effect,forastiere2021identification}.
    }
If we do this for each exposure, the resulting estimands may constitute a natural dose--response curve. Call the average outcome over such focal units an \emph{average focal expected outcome} (AFEO),
$$\afeo(\pi, f, y)\triangleq\avgpi{i\in \supp_f(\pi)}{\condexpectpi{\pi}{y_i(\vect{Z})}{f_i(\vect{Z})=1}},$$
where $\avgpi{i\in \supp_f(\pi)}{\cdot}$ denotes the conditional average over those units that have positive probability of being focal (a ``unit-averaging'') and $\condexpectpi{\pi}{\cdot}{\cdot}$ denotes the conditional expectation under policy $\pi$ (an ``assignment-averaging'').
This name reflects that this estimand has an inner expectation over treatment assignments, and then an outer average over units.
Call differences between AFEOs under different focal maps are \emph{focal contrasts} (e.g., a contrast in average outcomes between units with zero or one treated neighbor). Thus, comparisons between points in Figure~\ref{fig:aoe_vs_eao}a are focal contrasts.

Do these AFEOs and focal contrasts satisfy the two desiderata we have proposed?

To start with a favorable case, assume that the exposure map $d$ is correctly specified in the sense that it gives level sets in the potential outcome function; that is, if two treatment vectors $\vect{z} \neq \vect{z'}$ correspond to the same exposure, $d_i(\vect{z}) = d_i(\vect{z'})$, then they result in the same outcome for that unit, $y_i(\vect{z}) = y_i(\vect{z'})$.
Under this assumption, AFEOs defined by each unique exposure only have one meaningful averaging step --- averaging over units --- as all the potential outcomes pooled in the inner expectation are identical.

\paragraph*{As averages of unit-level causal effects.}
With a correctly specified exposure map, focal contrasts are readily interpretable as averages of unit-level causal effects.
In particular, with technical conditions about units having positive probability of being focal, focal contrasts are unit-level causal in any policy space that includes all deterministic interventions (i.e. all possible treatment assignments; Corollary~\ref{thm:focal_contrast_correct_exposure}). That means that we can interpret them as averages of the effect of assigning each unit to some exposure rather than some other exposure.
Whether these focal contrasts are particularly meaningful nonetheless depends on the choice of the exposure map, as arbitrary unit-specific garblings of a correctly specified exposure map are likewise correctly specified.

\paragraph*{Insufficiency for policy choice.}
We might expect that correctly specified exposure maps also make AFEOs useful for policy choice.
Empirical researchers often estimate AFEOs and focal contrasts and then draw intuitive conclusions about policies for assigning treatment based on those quantities. For example, \citet{cai2015insurance} use analogs of AFEOs estimated via regression to suggest the utility of social norms interventions, and \citet{forastiere2024causal} examine the dose--response curves given by AFEOs to suggest what the optimal agricultural subsidy regime is accounting for spillovers. 
However, as we show here, they are often not sufficient for policy choice; that is, this step from AFEOs to policy choices, while intuitive, is typically not formally justified.

To begin with a simple example, consider a setting where each unit's outcome only depends on whether 0, 1, or at least 2 of its neighbors are treated; it is unaffected by its own treatment. Thus, the true exposure mapping for each unit just has three levels, $d_i: \booln \rightarrow \{0, 1, 2+\}$.
The AFEOs are then $\overline{y}(0), \overline{y}(1),$ and $ \overline{y}(2+))$, where
$
\overline{y}(e) := \avgpi{i\in[n]}{y_i(e)}
$
and we abuse the notation to denote potential outcomes under exposure $e$. Differences between these can be seen as spillover effect (i.e. exogeneous peer effects). Let us imagine that average outcomes are maximized with a single treated neighbor, i.e. $\overline{y}(1) > \max(\overline{y}(0), \overline{y}(2+))$.

These quantities are a limited guide to policy. First, simply observing that the AFEOs are maximized at exactly one treated neighbor does not provide a way to achieve that average outcome. We typically cannot select a policy that directly corresponds to this average potential outcome: for many graphs, it will not be possible to assign treatments so that all units have exactly one treated neighbor.
Thus, if these AFEOs are to be relevant to policy choice, only some of them will be relevant (i.e. for the choice of all-or-none-treated policy, $\overline{y}(0)$ and $\overline{y}(2+)$ are relevant), or they will need to be combined to choose among policies that treat some units at random --- though their usefulness here may also be limited, as we now describe.

Consider a network given by Figure \ref{fig:example_graph}, or a network consisting of many copies of this graph. Say we are choosing between two (homogeneous Bernoulli) policies for treating units: one policy treats each unit with probability $\pi$, while the other treats each unit with probability $\pi' > \pi$. Can we use the AFEOs to make the optimal choice? No. These estimands have ``compressed'' the potential outcomes in a way that is not very useful for this purpose. The same three averages can arise from tables of potential outcomes that have very different implications for policy choice.\footnote{For example, say all outcomes are zero, except in some cases when a unit has exactly one treated neighbor, it may have a positive outcome (i.e. $y_i(e) = 0$ for all $e \neq 1$). In one setting, it is the higher-degree units who are helped by having one treated neighbors; in another, it is the lower-degree units who are so helped. Both of these settings can yield identical AFEOs. However, for a decision maker aiming to maximize utilitarian welfare, the optimal choice of $\pi$ or $\pi'$ is different across these settings.
}
Attempts to use these averages for policy choice, such as through a plug-in estimator of the expected average outcome under each policy, will be misleading. Here one problem is that we have discarded information about how likely different units are to receive different exposures under relevant policy changes.
In the biclique example of Figure \ref{fig:example_graph}, there is no Bernoulli policy with heterogeneous probabilities for units with different degrees (other than treating no or all units) that induces homogeneous exposure; see Figure \ref{fig:bicliquebern}.

More generally, the expected utilitarian welfare under a policy $\pi$ is identified by AFEOs of a correctly specified exposure map if and only if the exposure distribution under $\pi$ is unit-homogeneous --- that is, all units have the same probability distribution over exposures, $\probpi{\pi}{d_i(\vect{Z}) = e}$ independent of $i$ (Proposition~\ref{thm:eao_identification} and Corollary~\ref{thm:correctly_specified_offpolicy}).
On irregular graphs, many policies of typical interest will not produce such homogeneous exposure; in fact, no such non-trivial policy may exist.
When the exposure distribution is not unit-homogeneous, the welfare is only partially identified from the AFEOs --- and these bounds can be wide. To illustrate this, we reanalyze insurance adoption decisions in the field experiment of \citet{cai2015insurance}. Figure~\ref{fig:cai_partialidentification} shows both estimated AFEOs by number of treated neighbors and the bounds these imply for the welfare under Bernoulli policies with different probabilities of treatment $\pi$. These bounds collapse to a single point when $\pi = 0$, since this policy induces homogeneous exposure (i.e., all units have zero treated neighbors), but they are wide for other values of $\pi$. In this case, because some units do not have 2 or more neighbors, the bounds do not collapse to a point when $\pi = 1$, as this still induces a heterogeneous exposure distribution. 

Another view of the challenges here is available by considering the causal relationships between policies, treatments, exposures, and outcomes when the exposure mapping is correctly specified --- represented by the directed acyclic graph in Figure \ref{fig:dag}. AFEOs summarize how outcomes vary with exposures $D_i$, but a decision maker chooses a policy $\pi$, which only affects $D_i$ indirectly through $\vect{Z}$. It can be easy to end up considering interventions on the exposures $D_i$ that cannot be induced by any distribution for $\vect{Z}$ (that is, any choice of policy $\pi$).
Or, even if such a policy exists that induces a particular distribution on $D_i$, that policy may not be of interest at all (e.g., only simple policies may be feasible).\footnote{This is analogous to arguments, in the literature on mediation, that estimands involving cross-world counterfactuals are not relevant to policy \citep[e.g.][]{naimi2014mediation}.}

\paragraph*{With a misspecified exposure map.}
Researchers have been concerned about the consequences of misspecifying the exposure map, so it is worth understanding how both desiderata fare in this case.\footnote{One concern is that if a unit's outcomes are affected by their own treatment and neighbors' outcomes, this leads to potentially global dependence \citep[e.g.][]{manski2013identification,leung_discussion_2024, eckles2017design}.}
The present analysis is readily applicable to this case, as the inner averaging step for AFEOs is an expectation over potential outcomes grouped together by the exposure map, whether or not they form a level set.

Focal contrasts defined by misspecified exposure maps will generally remain averages of unit-level causal effects, though it can be difficult to understand exactly what they are effects \emph{of} \citep[cf.][\S S2]{savje2023exposure}. That they are causally interpretable has been suggested by work labeling them ``exposure effects'' \citep{savje2023exposure} or ``average distributional shift effects'' \citep{hudgens2008toward}. Focal contrasts are generally unit-level causal in some policy space (Proposition \ref{thm:focal_contrast_causal}), but this policy space may not be interpretable or have a straightforward relationship to how exposures are defined and labeled.
This adds to prior characterizations of specific cases where focal contrasts remain interpretable even with misspecification \citep{savje2021ateunknown}.

Regarding policy choice, if AFEOs are insufficient even when the exposure map is correctly specified, misspecification only compounds the problem.\footnote{In noting that misspecified exposure mappings can be irrelevant to policy choice, \citet{auerbach_discussion_2024} use an example in which the correctly-specified mapping is sufficient, but this depends on other characteristics of that example.}
Avoiding misspecification can motivate researchers to use a more granular exposure map; however, in addition to statistical estimation and inference challenges, making an already correctly specified exposure map more granular can sometimes result in making an exposure distribution that was homogeneous under a policy $\pi$ inhomogeneous, and thus making the welfare no longer point identified (Corollary \ref{thm:eao_degenerate}).\footnote{
    One specific example is that exposure maps giving the fraction of neighbors who are treated (i.e. $d_i(\vect{Z}) = \avgpi{j: A_{ij} = 1}{Z_j}$, where $A$ is the adjacency matrix) yield homogeneous exposure under global treatment and global control. On the other hand, in irregular graphs, this is not true under global treatment for an exposure map giving the number of treated neighbors.
}

To summarize, these widely-used assignments-then-units averages are often interpretable as averages of unit-level effects, though some of the clarity here depends on correctly specifying an exposure model. But, even in some of the best cases --- that is, even when the exposure map is correctly specified and these quantities have a clear causal interpretation --- they are often a very limited guide to policy choice. To be clear, the methodological literature working with AFEOs has typically not promised policy relevance, but empirical researchers have nonetheless often interpreted these quantities as relevant to policy, as is deceptively intuitive.

\section*{Averaging over units, then over assignments}
We can also represent some estimands as first averaging over units and then over possible treatment assignments. In the simplest case, we can, for each possible treatment assignment, average the potential outcomes of all units; then we marginalize over those average potential outcomes according to the probability of each under a policy of interest. This \emph{expected average outcome} (EAO) directly addresses the question of what happens in expectation under that policy. And, if the decision maker's utility is linear in the average outcome, it is sufficient for policy choice. In the context of causal inference in networks, the EAO (or the difference in EAO under two policies) has been studied in some recent work \citep[e.g.,][]{chin2022evaluating,viviano2025policy}.
One observation is simply that EAOs under policies of interest should continue to receive attention from methodologists, and empirical researchers interested in informing policy may wish to estimate EAOs.

It is also possible to define other quantities less familiar in the causal inference literature. 
To again consider something like a spillover effect, we can, for each possible treatment assignment $\vect{z} \in \{0, 1\}^n$, summarize the potential outcomes of all units that had different number of treated neighbors in that assignment (Figure \ref{fig:overview}, lower left). Then we can average these entries to produce average outcomes for units with 0, 1, or at least 2 treated neighbors under some distribution for $\vect{Z}$, some policy of interest $\pi$. We call these \emph{expected focal average outcomes} (EFAOs),
$$
\efao(\pi, f, y)\triangleq\condexpectpi{\pi}{\avgpi{f_i(\vect{Z})=1}{y_i(\vect{Z})}}{f(\vect{Z})\ne\vect{0}},
$$
where $\condexpectpi{\pi}{\cdot}{\cdot}$ denotes the conditional expectation over $\vect{Z}$ w.r.t. policy $\pi$ (an ``assignment-averaging'') and $\avg{\cdot}_{f_i(\vect{Z})=1}$ denotes the conditional average over focal units, a ``unit-averaging''.
EFAOs and AFEOs are thus named with the same words in different orders, reflecting that they perform the same two averaging steps (over units and over treatment assignments) but in the opposite order. Such EFAOs --- defined by the number of treated neighbors --- would then answer the question of what the average outcome of units with, e.g., exactly one treated neighbor is under a policy $\pi$.

Do these EFAOs and contrasts between them for different policies or different focal maps satisfy the two desiderata we have proposed?

\paragraph*{Generally not unit-level causal.}
Contrasts of EFAOs for two different policies, $\efao(\pi, f, y) - \efao(\pi', f, y)$, do not generally have an interpretation as an average of unit-level causal effects.
There are some special cases where an EFAO contrast coincides, exactly or asymptotically, with an intuitively related quantity represented via the other route (Proposition \ref{thm:policy_contrast_causal}). For example, this comparison of average outcomes for treated and untreated units is the focal contrast of the unit's own treatment when $\pi$ is a completely randomized (i.e. $m$-out-of-$n$) design, and it is the global average treatment effect \citep[GATE;][]{ugander2013graph} when $\pi$ is an all-or-none-treated policy.
These are all cases where the policy induces homogeneous probabilities of each exposure. 

Outside of such special cases, EFAO contrasts are not unit-level causal. Under a heterogeneous Bernoulli policy (e.g., treating high degree nodes with higher probability), for example, this kind of one-treated-neighbors vs. no-treated-neighbors comparison can be large even with no treatment effects whatsoever. More generally, EFAO contrasts are only unit-level causal in all policy spaces if the focal units are selected determistically (i.e. the focal map is invariant in $\vect{z}$; see Proposition \ref{thm:policy_contrast_det_focal_causal}). That is, they only have a unit-level causal interpretation when they are not really summaries of outcomes for units with different exposures, but rather just summaries of outcomes for the whole population (i.e. the EAO) or a fixed subset of units.

\paragraph*{Other causal interpretations.}
If these quantities are often not unit-level causal, are they useful for causal reasoning?
We contend that these quantities are often nonetheless causally interpretable, but in a different way.
First, a contrast of these quantities between two policies $\pi$ and $\pi'$ is still a causal effect of policy on \emph{aggregate} outcomes. Perhaps $\pi$, compared with $\pi'$, increases what treated units' neighbors' outcomes are on average. These are causal quantities, describing how the policy chosen affects aggregate outcomes (e.g., what will outcomes of neighbors of treated units look like?), even if they are not averages of \emph{unit-level} effects.

\paragraph*{Sufficiency for policy choice.}
Even if these quantities are not unit-level causal, they may still be sufficient for policy choice.
First, if a decision maker is maximizing utilitarian welfare, then the EFAO with all units as focal (i.e. EAO) is sufficient for policy choice.
Second, decision makers are often evaluated, at least in part, based on myopic summaries (e.g., only looking at treated units) or naive comparisons (e.g., of treated and untreated units).
Thus, rational (if perhaps cynical) decision makers will have preferences that separately weigh outcomes for treated and untreated units.
They may prefer treated units to have good outcomes; or they may prefer treated units to not have outcomes that are \emph{too good}, in the case of social programs that they want to appear to be \emph{ex post} targeted to the less well-off.
Furthermore, in the context of causal inference in networks, they may have preferences about the outcomes of units with various exposures to treatment.
In a social network, a decision maker may have preferences about the outcomes of people within one hop of units selected for an intensive marketing intervention \citep[as in, e.g.,][]{cai2015insurance}.
While some of these preferences might arise in settings without interference, positing interference often means that we expect that units will continue to interact, such that unit $i$'s intermediate outcome should be interpreted in the context of aggregate outcomes as they will, e.g., be competing for scarce resources using the income it represents.

Thus, particularly in the presence of interference, it will be natural for decision makers to have complex aggregations of outcomes in mind when choosing among policies. 
This is represented in Figure \ref{fig:dag}, where exposures are potential inputs to welfare.

\section*{Conclusion}

How then should we quantify outcomes and effects of interventions in networks?
Decision makers need quantities that are relevant for policy choice. And the causal inference literature typically prefers quantities that are summaries of unit-level causal effects.
The expected average outcome~(EAO) is a quantity that coincides under both kinds of averaging we have described, under \emph{any} treatment policy, regardless of whether it induces different exposures homogeneously or not.\footnote{Thus we could equally call this the average expected outcome, as either ordering of unit-averaging and assignment-averaging yields the same quantity.}
Contrasts of EAOs are thus both policy-relevant --- since we marginalize over the treatment of all units under the same policy and therefore can compare this quantity across different policies --- and are readily interpretable as summaries of unit-level causal effects.
Rather than fixating on specific unit-level causal contrasts that are informative only about policies that are homogeneous in the exposures they induce, we argue that estimating EAOs --- for all policies within a larger relevant policy space --- will be more directly decision-relevant.
The EAO is the unique estimand for which both averaging routes coincide universally (Proposition~\ref{thm:commuting_maps_deterministic}), and that it is necessary and sufficient for policy choice under utilitarian welfare with arbitrary costs (Proposition~\ref{thm:efao_policychoice}).

We are not arguing for neglecting other estimands altogether. Researchers may often also be interested in AFEOs like those in Figure \ref{fig:aoe_vs_eao}a as part of basic questions in behavioral science. These quantities may play a central role in tests of theory-driven choices of exposure maps.
But, to the extent that researchers have more applied goals as well, these should be supplemented with other quantities that are sufficient for policy choice.

Some other estimands in the literature, which take forms not readily included in our characterization here, may be able to be combined to yield a difference in EAOs for pairs of policies while also each being interpretable as summaries of unit-level causal effects. For example, for comparing policies that treat $m$ versus $m-1$ units, we could decompose this difference in EAOs into a direct effect and an average (or total) effect of treating a random unit on all other units \citep{hu2022indirecteffects}. We encourage further attention to decompositions of EAOs and differences in EAOs that may provide additional insight into the mechanisms underlying aggregate consequences of a treatment policy.

\subsection*{Disclosure Statement}
The authors have no conflicts of interest to declare.

\subsection*{Acknowledgments}
SL was supported by a Schmidt Science Fellowship. Some ideas in this paper developed in part in response to the Symposium on Causality in Florence and the workshop on Causal Inference and Prediction for Network Data at the Banff International Research Station in 2024. We are grateful for comments from Daniel Nevo, Davide Viviano, and seminar participants at the Harvard Institute for Quantitative Social Science, Harvard Business School, and the Yale Institute for Foundations of Data Science.
 
\subsection*{Contributions}
SL and DE conducted the research and wrote the article.

\bibliography{references}

\begin{landscape}
\begin{figure}
    \centering
    
    \scalebox{0.78}{
\begin{tikzpicture}

  \node[draw, fill=blue!3, rounded corners, align=center] (science_table) {
    \textbf{\textsc{Table A}}\\
    \begin{tabular}{>{\centering\arraybackslash}p{1.4cm}||c|c|c|c|c|c|c|>{\columncolor{red!20}}c|c|c}
       \multirow{2}{*}{Unit $i$} & \multicolumn{9}{c}{Treated units $\set{j\in[n]}{Z_j=1}$=}\\\cline{2-11}
        & $\emptyset$ & $\{1\}$ & $\{2\}$ & $\cdots$ & $\{1,2\}$ & $\{1,3\}$ & $\cdots$ & $V$ & $\cdots$ & $[n]$\\
      \hline\hline
      1 & 4 & 3 & 4 & $\cdots$ & 4 & 2 & $\cdots$ & 3 & $\cdots$ & 4\\
      2 & 3 & 4 & 5 & $\cdots$ & 7 & 3 & $\cdots$ & 2 & $\cdots$ & 5\\
      $\vdots$ & $\vdots$ & $\vdots$ & $\vdots$ & $\vdots$ & $\vdots$ & $\vdots$ & $\vdots$ & $\vdots$ & $\vdots$ & $\vdots$\\
      $n$ & 5 & 2 & 5 & $\cdots$ & 2 & 2 & $\cdots$ & 5 & $\cdots$ & 8
    \end{tabular}
  };
  
  \node[draw, fill=green!3, rounded corners, right=5.5cm of science_table, align=center] (zavg_table) {
  \textbf{\textsc{Table B}}\\
  \begin{tabular}{>{\centering\arraybackslash}p{1.4cm}||>{\centering\arraybackslash}p{1.8cm}|>{\centering\arraybackslash}p{1.8cm}|>{\centering\arraybackslash}p{1.8cm}}
       \multirow{2}{*}{Unit $i$} & \multicolumn{3}{c}{Number of treated neighbors $D_i=$} \\\cline{2-4}
        & $0$ & $1$ & $2+$ \\
      \hline\hline
      1 & 5.2 & 4.1 & 4.8\\
      2 & 2.5 & 3.8 & 2.9\\
      $\vdots$ & $\vdots$ & $\vdots$ & $\vdots$\\
      $n$ & 7.3 & 5.0 & 6.1
    \end{tabular}
  };

  \node[draw, fill=green!3, rounded corners, below=1.5cm of zavg_table, align=center, xshift=-1.5cm] (zyavg_table) {
  \parbox{10cm}{\textbf{\textsc{Table C}: Average focal expected outcomes (AFEOs)}}\\
    \begin{tabular}{>{\centering\arraybackslash}p{1.4cm}||c|c|c}
        & \multicolumn{2}{c}{Estimands}\\\cline{2-4}
        & $\avgpi{i\in[n]}{\condexpectpi{\pi}{Y_i}{D_i=0}}$ & $\avgpi{i\in[n]}{\condexpectpi{\pi}{Y_i}{D_i=1}}$ & $\avgpi{i\in[n]}{\condexpectpi{\pi}{Y_i}{D_i=2+}}$\\
      \hline\hline
      Value & 4.2 & 4.5 & 4.9
    \end{tabular}
  };

  \node[draw, fill=green!3, rounded corners, below=5cm of science_table, align=center] (yavg_table) {
  \textbf{\textsc{Table D}}\\
    \begin{tabular}{>{\centering\arraybackslash}p{1.4cm}||c|c|c|c|c|c|c|>{\columncolor{red!20}}c|c|c}
       Units w/ & \multicolumn{9}{c}{Treated units $\set{j\in[n]}{Z_j=1}$=}\\\cline{2-11}
        $D_i=$ & $\emptyset$ & $\{1\}$ & $\{2\}$ & $\cdots$ & $\{1,2\}$ & $\{1,3\}$ & $\cdots$ & $V$ & $\cdots$ & $[n]$\\
      \hline\hline
      $0$ & 3.4 & 3.5 & 3.2 & $\cdots$ & 5.5 & 6.7 & $\cdots$ & 6.1 & $\cdots$ & \\
      $1$ &  & 3 & 5 & $\cdots$ & 5.5 & 4 & $\cdots$ & 7.2 & $\cdots$ & 9.6 \\
      $2+$ &  &  &  &  & 6.2 & 5.1 & $\cdots$ & 6.7 & $\cdots$ & 9.6
    \end{tabular}
  };

  \node[draw, fill=green!3, rounded corners, right=4cm of yavg_table, align=center] (yzavg_table) {
  \parbox{5cm}{\centering\textbf{\textsc{Table E}: Expected focal average outcomes (EFAOs)}}\\
  \begin{tabular}{c||c}
       \multirow{2}{*}{Estimands} & \multirow{2}{*}{Value}\\
     \\
      \hline\hline
      $\condexpectpi{\pi}{\avgpi{D_i=0}{Y_i}}{\exists i: D_i=0}$ & 4.6 \\      
      $\condexpectpi{\pi}{\avgpi{D_i=1}{Y_i}}{\exists i: D_i=1}$  & 5.8 \\
      $\condexpectpi{\pi}{\avgpi{D_i=2+}{Y_i}}{\exists i: D_i=2+}$ & 5.2
    \end{tabular}
  };

  \draw[->, thick] (science_table) -- node [midway, above, align=center] {Marginalize outcomes\\ over treatments under} node [midway, below, align=center] {unit-exposure-conditional policies} (zavg_table);
  \draw[->, thick] (science_table) -- node [midway, left] {Average outcomes} node [midway, right] {of treatment-conditional units} (yavg_table);
   \draw[->, thick] ([xshift=-1.5cm]zavg_table.south) -- node [midway, left] {Averaging outcomes} node [midway, right] {of all $n$ units} (zyavg_table.north);
   \draw[->, thick] (yavg_table) -- node [midway, above, align=center] {Marginalize outcomes \\over treatments} node [midway, below, align=center] {under policy $\pi$} (yzavg_table);
    \draw[<->, thick, gray] (yzavg_table.east) to[out=0, in=-25] node [midway, right, align=center, black] {generally not equal} ([xshift=2.5cm]zyavg_table.south);
\end{tikzpicture}
}
    \caption{The complete set of potential outcome functions $y_i(\vect{z})$ for every unit $i\in[n]$ is encoded in the unknown ``science'' table (in blue), where each row is a unit $i$, and each column is a set of treated units $\set{j\in[n]}{Z_j=1}$. We only observe one column (in red) of the table: $Y_i \triangleq y_i(\vect{Z})$ for the realized treatment vector $\vect{Z}\in\booln,\vect{Z}\sim\pi$, $V\triangleq\set{j\in[n]}{Z_j=1}$. We can ``compress'' this table into interpretable estimands by moving in one of two directions: (a) integrating over the columns, i.e. marginalizing over the treatments under unit-specific policies $\pi_i$ (move right), or (b) integrating over the rows, i.e. averaging the outcomes over treatment-specific node sets (move down). These averages can be executed in succession --- a form of ``double averaging'' --- and the resulting estimands will be typically sensitive to their ordering. In other words, the averaging operations generally do not commute; see Proposition \ref{thm:commuting_fixedcardinality}. Here, the averaging is done with respect to the ``exposure map'' $D_i \triangleq d_i(\vect{Z})$ of the number of treated neighbors a unit has.}
    \label{fig:overview}
\end{figure}
\end{landscape}

\begin{figure}
\begin{center}
\begin{subfigure}[t]{0.3\textwidth}
\centering
\drawbiclique{2}{3}
\caption{}
\label{fig:example_graph}
\end{subfigure}
\hfill
\begin{subfigure}[t]{0.65\textwidth}
\centering
\begin{tikzpicture}[
  node distance=1.5cm,
  every node/.style={font=\Large},
  >=Stealth
]
  \node (Pi) {$\pi$};
  \node[right=of Pi]  (Z)  {$\vect{Z}$};
  \node[right=of Z]   (Di) {$D_i$};
  \node[below=of Z, yshift=0.5cm] (A) {$\mat{A}$};
  \node[right=of Di]  (Yi) {$Y_i$};
  \node[right=of Yi]  (W) {$W$};
  \draw[->] (Pi) -- (Z);
  \draw[->] (Z)  -- (Di);
  \draw[->] (Di) -- (Yi);
  \draw[->] (Yi) -- (W);
  \draw[->] (A) -- (Di);
  \draw[->, bend left=40] (Di) to (W);
  \begin{scope}[on background layer]
    \node[draw,rounded corners,fit=(Di)(Yi),inner sep=8pt,
          label=below right:{$i\in\{1,\ldots,n\}$}] {};
  \end{scope}
\end{tikzpicture}
\caption{}
\label{fig:dag}
\end{subfigure}\\
\begin{subfigure}[t]{\textwidth}
    \centering
    \begin{tikzpicture}[declare function={binom(\k,\n,\p)=\n!/(\k!*(\n-\k)!)*\p^\k*(1-\p)^(\n-\k);}]
            \begin{groupplot}[group style={group size=3 by 1, horizontal sep=.5cm},
            width=5.5cm, height=4cm, xmin=0, xmax=1]
            \nextgroupplot[title={Exposure $d=0$}, ylabel={$\probpi{\pi}{D_i=d}$}, ylabel style={yshift=-0.3cm}, xlabel={Probability of treatment $\pi$},legend style={draw=none, fill=none, font=\small}]
            \addplot[domain=0:1, smooth, ultra thick, red]{binom(0, 2, x)};
            \addplot[domain=0:1, smooth, ultra thick, blue]{binom(0, 3, x)};
            \legend{Degree 2, Degree 3} 
            \nextgroupplot[title={Exposure $d=1$}, yticklabels={}, xlabel={Probability of treatment $\pi$}]
            \addplot[domain=0:1, smooth, ultra thick, red]{binom(1, 2, x)};
            \addplot[domain=0:1, smooth, ultra thick, blue]{binom(1, 3, x)};
            \nextgroupplot[title={Exposure $d=2+$}, yticklabels={}, xlabel={Probability of treatment $\pi$}]
            \addplot[domain=0:1, smooth, ultra thick, red]{binom(2, 2, x)};
            \addplot[domain=0:1, smooth, ultra thick, blue]{binom(2, 3, x)+binom(3, 3, x)};
            \end{groupplot}
            \end{tikzpicture}
    \caption{}
    \label{fig:bicliquebern}
\end{subfigure}
\begin{subfigure}{\textwidth}
    \centering
    \includegraphics[width=0.8\linewidth]{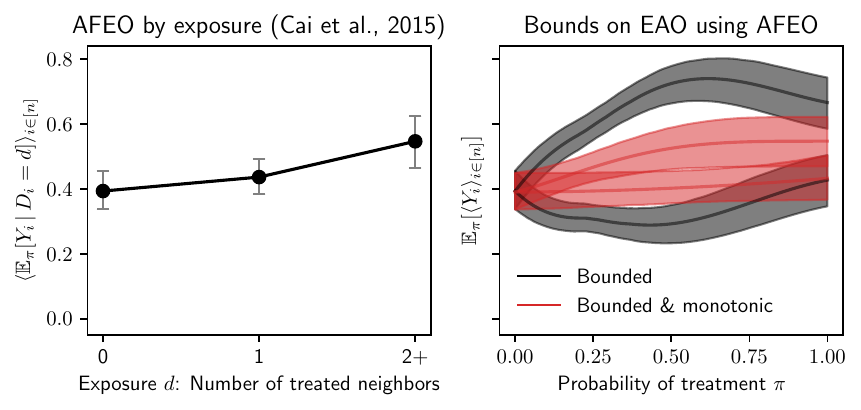}
    \caption{}
    \label{fig:cai_partialidentification}
\end{subfigure}
\end{center}
\vspace{-0.7cm}
\caption{
\emph{(a)} A biclique $K_{2,3}$.
\emph{(b)} Directed acyclic graph representing the causal relationships in a setting with interference according to a correctly specified exposure map. The policy $\pi$ produces the treatment assignment vector $\vect{Z}$, which determines, in a unit-specific way, each unit's exposure $D_i$ alongside the full network structure encoded in the adjacency matrix $\mat{A}$. If the exposure mapping is correctly specified, then $\vect{Z}$ only affects $Y_i$ via $D_i$. The decision maker is interested in some aggregate quantity (e.g., welfare) given by $W$.
\emph{(c)} There are no non-trivial (heterogeneous) Bernoulli policies with varied probabilities of treatment for degree-2 and degree-3 units that induce equal probability of exposure for the exposure mapping of the number of treated neighbors $\{0, 1, 2+\}$, except for the all-or-none-treated policies; that is, there is no other common points where the these curves meet in all three panels. So there is no such non-trivial policy for which the average outcomes by exposure are jointly relevant.
\emph{(d)} When probabilities of exposure are not homogeneous across units, we can only partially identify the expected average outcomes from the average outcomes by exposure --- even when the exposure map is correctly specified; see Corollary \ref{thm:correctly_specified_offpolicy} --- as shown here using data from \cite{cai2015insurance}. AFEOs by exposure (left) only partially identify the EAO curve (right). Lines indicate bounds on the EAO, with red lines being the bounds when outcomes are assumed to be monotonic under exposure levels. Error bars and bands are 95\% confidence intervals.
}
\end{figure}

\clearpage

\section*{Appendix}

\begin{table}[ht!]
    \centering
    \begin{tabular}{l|p{12cm}}
        Result & Description \\\hline
        Proposition \ref{thm:commuting_maps_deterministic} & AFEO and EFAO coincide $\forall\mathcal{P}_n$ iff $f$ is deterministic \\
        Proposition \ref{thm:focal_contrast_causal} & AFEO contrasts$^\dagger$ are unit-level causal in some $\mathcal{P}_n$ \\
        Corollary \ref{thm:focal_contrast_correct_exposure} & AFEO contrasts$^\dagger$ from a correctly specified exposure map are unit-level causal in any $\mathcal{P}_n$ with all deterministic interventions \\
        Proposition \ref{thm:policy_contrast_causal} & EFAO contrasts$^\dagger$ are unit-level causal in some $\mathcal{P}_n$ iff the harmonic version of $\mathcal{P}_n$ is projective-marginally equivalent w.r.t. $f$\\
        Proposition \ref{thm:focal_contrast_universal_causal} & AFEO contrasts$^\dagger$ are \emph{not} unit-level causal $\forall\mathcal{P}_n$\\
        Proposition \ref{thm:policy_contrast_det_focal_causal} & EFAO contrasts$^\dagger$ are unit-level causal $\forall\mathcal{P}_n$ iff $f$ is deterministic\\
        Proposition \ref{thm:efao_policychoice} & Estimand $T$ is sufficient for policy choice $\forall\mathcal{P}_n$ w.r.t. utilitarian welfare with arbitrary costs iff $T$ functionally identifies EFAO up to a constant\\
        Corollary \ref{thm:correctly_specified_offpolicy} & AOE under a correct exposure map identifies EAO at policy $\pi$ iff the exposure distribution under $\pi$ is unit-homogeneous\\\hline
        &$^\dagger${\footnotesize that are support-aligned and non-degenerate}
    \end{tabular}
    \caption{A summary of key results used in the main text.}
    \label{tab:summary_results}
\end{table}

\begin{definition}[Potential outcome]\label{def:potential_outcome}
Outcomes are elements in a set $\mathcal{B}\subset\real$ with $\mathcal{B}$ bounded. The potential outcome function space $\mathcal{Y}_n$ is a subspace of bounded pseudo-Boolean functions over $\booln$: $$\mathcal{Y}_n\subseteq\set{y:\booln\to\mathcal{B}^n}{\mathcal{B}\text{ is bounded.}}.$$
An element of this space $y\in\mathcal{Y}_n$, called a potential outcome function, yields the outcomes of units in a population of $n$ units under all possible treatments. $\mathcal{Y}_n$ is said to be nontrivial if $\abs{\mathcal{B}}\ge 2$.
\end{definition}

\begin{remark}
    We use the notation $y_i$ to refer to the $i^\text{th}$ component of a vector-valued function $y$; if $y$ is the potential outcome function of the entire population then $y_i$ is that of unit $i$.
\end{remark}

\begin{definition}[Policy]\label{def:policy_space}
    Policy space $\mathcal{P}_n$ is a subspace of probability distributions over $\booln$: $$\mathcal{P}_n\subseteq\set{p:\booln\to[0,1]}{\sum_{\vect{z}\in\booln}p(\vect{z})=1}.$$
    An element of this space $\pi\in\mathcal{P}_n$, called a policy, encodes the distribution of treatments in a population of $n$ units. $\mathcal{P}_n$ is said to be nontrivial if $\abs{\mathcal{P}_n}\ge 2$.
\end{definition}

\begin{definition}[Focal map]\label{def:focal_units} Focal space $\mathcal{F}_n$ is a subspace of Boolean transformations over $\booln$: $$\mathcal{F}_n\subseteq\set{f:\booln\to\booln}{\exists\vect{z}\in\booln:f(\vect{z})\ne\vect{0}}.$$
An element of this space $f\in\mathcal{F}_n$, called a focal map, indicates which units' outcomes are of interest with respect to the treated units of the population.
\end{definition}

\begin{remark}
    A focal map can be seen as a special kind of exposure map \citep{aronow2017interference} that indicates which units receive one exposure. For an exposure map that gives the number of treated neighbors, then for the exposure with exactly $d$ treated neighbors, if the interference network is encoded by the Boolean matrix $\mat{A}$, then the corresponding focal map is $f_d(\vect{z})=\ones{\mat{A}\vect{z}=d}$. See also Definition \ref{def:exposure_map}.
\end{remark}

\begin{definition}[Deterministic focal map]\label{def:deterministic}
    A focal map $f$ (Definition \ref{def:focal_units}) is said to be deterministic if $\forall\vect{z},\vect{z}'\in\booln: f(\vect{z})=f(\vect{z}')$.    
\end{definition}

\begin{definition}[Positive policy w.r.t. focal map]\label{def:positivity} 
    Define the support of a focal map $f$ (Definition \ref{def:focal_units}) under policy $\pi$ (Definition \ref{def:policy_space}) as $$\supp_f(\pi)\triangleq\set{i\in[n]}{\expectpi{\pi}{f_i(\vect{Z})}>0}.$$
    Then $\pi$ is said to be positive w.r.t. $f$ if $\supp_f(\pi)\ne\emptyset$, and \emph{globally} positive if $\supp_f(\pi)=[n]$.
\end{definition}

\begin{remark}
    We use the notation ``$\mathsf{X}$ has property $\mathsf{P}$ w.r.t. policy space $\mathcal{P}_n$'' iff ``$\forall\pi\in\mathcal{P}_n: \mathsf{X}$ has property $\mathsf{P}$ w.r.t. policy $\pi$.'' Similarly, we use the notation ``$\mathsf{X}$ has property $\mathsf{P}$ w.r.t. focal space $\mathcal{F}_n$'' iff ``$\forall f\in\mathcal{F}_n: \mathsf{X}$ has property $\mathsf{P}$ w.r.t. focal map $f$.''
\end{remark}

\begin{definition}[Homogeneous focal map]\label{def:homogeneous_focal}
    A focal map $f$ (Definition \ref{def:focal_units}) is said to be homogeneous under policy $\pi$ (Definition \ref{def:policy_space}) if $\exists c\in[0,1]$ such that $\forall i\in\supp_f(\pi): \expectpi{\pi}{f_i(\vect{Z})}=c$. $f$ is said to be \emph{conditionally} homogeneous under positive policy $\pi$ (Definition \ref{def:positivity}) if $\exists c\in(0,1]$ such that $\forall i\in\supp_f(\pi): \condexpectpi{\pi}{f_i(\vect{Z})}{f(\vect{Z})\ne\vect{0}}=c$.
\end{definition}

\begin{proposition}\label{thm:cond_homo}
    A focal map $f$ is conditionally homogeneous under positive policy $\pi$ (Definitions \ref{def:positivity}, \ref{def:homogeneous_focal}) iff it is homogeneous under $\pi$.
\end{proposition}

\begin{proof}
    \sloppy For a positive policy $\pi$ we have $\expectpi{\pi}{f_i(\vect{Z})}=\probpi{\pi}{f(\vect{Z})\ne\vect{0}}\condexpectpi{\pi}{f_i(\vect{Z})}{f(\vect{Z})\ne\vect{0}}$ and $\probpi{\pi}{f(\vect{Z})\ne\vect{0}}\triangleq c_{f,\pi}>0$. Therefore, $f$ is conditionally homogeneous under $\pi$ iff it is homogeneous under $\pi$.
\end{proof}

\begin{definition}[Fixed-cardinality focal map]\label{def:fixedcardinality_focal}
    A focal map $f$ (Definition \ref{def:focal_units}) is said to be of fixed cardinality under policy $\pi$ (Definition \ref{def:policy_space}) if $\variancepi{\pi}{\abs{f(\vect{Z})}}=0$. $f$ is said to be of \emph{conditionally} fixed cardinality under positive policy $\pi$ (Definition \ref{def:positivity}) if $\condvariancepi{\pi}{\abs{f(\vect{Z})}}{f(\vect{Z})\ne\vect{0}}=0$.
\end{definition}

\begin{remark}
    By the law of total variance, if $f$ is of fixed cardinality under positive policy $\pi$ then it is of conditionally fixed cardinality under $\pi$, but the inverse need not be true.
\end{remark}

\begin{definition}[Focal map equivalence]\label{def:equivalence} 
    Define the marginal-difference support of a pair of focal maps $f,f'$ (Definition \ref{def:focal_units}) under policy $\pi$ (Definition \ref{def:policy_space}) as those units with that are sometimes focal under one map but not the other, i.e. 
    $$
    \diffsupp{f}{f'}(\pi)\triangleq\set{i\in[n]}{\expectpi{\pi}{(f_i(\vect{Z})-f_i'(\vect{Z}))^2}>0}.
    $$
    Then $f_i\stackrel{\pi}{\equiv}f_i'\iff i\notin\diffsupp{f}{f'}(\pi)$ and $f\stackrel{\pi}{\equiv}f'\iff\diffsupp{f}{f'}(\pi)=\emptyset$.
\end{definition}

\begin{definition}[Policy equivalence]\label{def:equivalence_policy} 
    Define the marginal-difference support of a focal map (Definition \ref{def:focal_units}) under a pair of policies $\pi,\pi'$ (Definition \ref{def:policy_space}) as those units that can be focal under one policy but not the other, or whose conditional law when focal differs across policies, i.e.
    $$
    \diffsupppi{f}(\pi,\pi')\triangleq\set{i\in [n]}{i\in\supp_f(\pi)\Delta\supp_f(\pi') \text{ or }\pi_i\ne\pi_i'},
    $$
    where $\pi_i$ is the conditional law when $f_i(\vect{Z})=1$.
    Then $\pi_i\stackrel{f}{\equiv}\pi_i'\iff i\notin\diffsupppi{f}(\pi,\pi')$ and $\pi\stackrel{f}{\equiv}\pi'\iff\diffsupppi{f}(\pi,\pi')=\emptyset$.
\end{definition}

\begin{remark}
    If $f$ is deterministic then $\pi\stackrel{f}{\equiv}\pi'\iff\pi=\pi'$.
\end{remark}

\begin{definition}[Estimand]\label{def:estimand}
    An estimand is a functional $\theta:\mathcal{P}_n\times\mathcal{F}_n\times\mathcal{Y}_n\to\real$ on the joint law of outcomes $y(\vect{Z})$ (Definition \ref{def:potential_outcome}) and focal vector $f(\vect{Z})$ (Definition \ref{def:focal_units}) under policy $\pi$ (Definition \ref{def:policy_space}): $$\theta(\pi,f,y)=\Phi(\probpi{\pi}{y(\vect{Z}), f(\vect{Z})}),$$
    and a space of estimands is denoted by $\Theta_n$. Naturally, a function of estimands (independent of $y$) is also considered to be an estimand.
\end{definition}

\begin{definition}[Contrast]\label{def:contrast}
    For a fixed potential outcome function $y$, a contrast is a difference in two estimands $\theta(\pi,f,y)-\theta'(\pi',f',y)$ (Definition \ref{def:estimand}). It is said to be support-aligned if $\supp_f(\pi)=\supp_{f'}(\pi')$ (Definition \ref{def:positivity}).
    
\end{definition}

\begin{definition}[Expected focal average outcome (EFAO)] \label{def:efao} 
    Consider a population of $n$ units with treatment vector $\vect{Z}\in\{0,1\}^n$ realized under policy $\pi$ and outcomes $y(\vect{Z})$ (Definition \ref{def:potential_outcome}). Then the expected focal average outcome (EFAO) under $\pi$ over focal map $f$, where $\pi$ is positive w.r.t. $f$ (Definition \ref{def:positivity}), is given by:
    $$\efao(\pi, f, y)\triangleq\condexpectpi{\pi}{\avgpi{f_i(\vect{Z})=1}{y_i(\vect{Z})}}{f(\vect{Z})\ne\vect{0}},$$
    where $\condexpectpi{\pi}{\cdot}{\cdot}$ denotes the conditional expectation over $\vect{Z}$ w.r.t. policy $\pi$, a ``treatment-averaging'', and $\avg{g_i}_{h_i}\triangleq\frac{\sum_{h_i\text{ is true}}{g_i}}{\sum_{h_i\text{ is true}}1}$ denotes the conditional average over focal units, a ``unit-averaging''.
\end{definition}

\begin{definition}[Average focal expected outcome (AFEO)] \label{def:afeo} 
    Consider a population of $n$ units with treatment vector $\vect{Z}\in\{0,1\}^n$ realized under policy $\pi$ and outcomes $y(\vect{Z})$ (Definition \ref{def:potential_outcome}). Then the average focal expected outcome (AFEO) over a focal map $f$ under $\pi$, where $\pi$ is positive w.r.t. $f$ (Definition \ref{def:positivity}), is given by:
    $$\afeo(\pi, f, y)\triangleq\avgpi{i\in \supp_f(\pi)}{\condexpectpi{\pi}{y_i(\vect{Z})}{f_i(\vect{Z})=1}},$$
    where $\avgpi{i\in \supp_f(\pi)}{\cdot}$ denotes the conditional average over focal units in $\supp_f(\pi)$, a ``unit-averaging'', and $\condexpectpi{\pi}{\cdot}{\cdot}$ denotes the conditional expectation w.r.t. policy $\pi$, a ``treatment-averaging''.
\end{definition}

\begin{remark}
    Both EFAO and AFEO perform unit-averaging and treatment-averaging --- a form of double-averaging --- but in the opposite order.
\end{remark}

\begin{definition}[Commuting focal map]\label{def:commuting} A focal map $f$ is said to be commuting for a policy space $\mathcal{P}_n$ if $\mathcal{P}_n$ is positive w.r.t. $f$ (Definition \ref{def:positivity}) and, for any fixed potential outcome function $y$, unit-averaging and treatment-averaging commute w.r.t. $\mathcal{P}_n$: $$\forall\pi\in\mathcal{P}_n:\afeo(\pi,f,y)=\efao(\pi,f,y).$$
It is said to be \emph{universally} commuting if it is commuting for any nonempty $\mathcal{P}_n$.
\end{definition}

\begin{remark}
    If a focal map maps to $\vect{0}$ then there exists a policy that is \emph{not} positive w.r.t. that focal map, rendering the averaging as undefined and implying that such focal maps cannot be \emph{universally} commuting. However, they might commute in restricted positive policy spaces, as we consider next.
\end{remark}

\begin{proposition}[Focal maps are commuting iff they are homogeneous and of conditionally fixed cardinality]\label{thm:commuting_fixedcardinality}
     A focal map $f$ is commuting for positive policy space $\mathcal{P}_n$ (Definition \ref{def:commuting}) iff it is homogeneous (Definition \ref{def:homogeneous_focal}) and of conditionally fixed cardinality (Definition \ref{def:fixedcardinality_focal}) under $\mathcal{P}_n$.
\end{proposition}

\begin{proof}
    From Definition \ref{def:efao},
    \begin{align*}
        \efao(\pi,f,y)&=\condexpectpi{\pi}{\frac{\sum y_i(\vect{Z})\,f_i(\vect{Z})}{\abs{f(\vect{Z})}}}{f(\vect{Z})\ne\vect{0}}\\&=\condexpectpi{\pi}{\condexpectpi{\pi}{\frac{\sum y_i(\vect{Z})\,f_i(\vect{Z})}{k}}{\abs{f(\vect{Z})}=k}}{f(\vect{Z})\ne\vect{0}}\\&=\condexpectpi{\pi}{k^{-1}\sum\condexpectpi{\pi}{y_i(\vect{Z})\,f_i(\vect{Z})}{\abs{f(\vect{Z})}=k}}{f(\vect{Z})\ne\vect{0}}\\&=\sum\condexpectpi{\pi}{k^{-1}\condexpectpi{\pi}{y_i(\vect{Z})\,f_i(\vect{Z})}{\abs{f(\vect{Z})}=k}}{f(\vect{Z})\ne\vect{0}}\\&=C_\pi^{-1}\sum\expectpi{\widetilde{\pi}}{f_i(\vect{Z})}\expectpi{\widetilde{\pi}_i}{y_i(\vect{Z})},
    \end{align*}
    where we use linearity of expectation, sums are over $i\in\supp_f(\pi)$, and $\widetilde{\pi}$ is the following probability law: $$\probpi{\widetilde{\pi}}{\cdot}\triangleq C_\pi \sum_{k=1}^\infty k^{-1}\condprobpi{\pi}{\cdot}{\abs{f(\vect{Z})}=k}\condprobpi{\pi}{\abs{f(\vect{Z})}=k}{f(\vect{Z})\ne\vect{0}},$$
    where $C_\pi\triangleq\left(\sum_{k=1}^\infty k^{-1}\condprobpi{\pi}{\abs{f(\vect{Z})}=k}{f(\vect{Z})\ne\vect{0}}\right)^{-1}$ is the conditional harmonic mean of the number of focal nodes --- conditional on them being positive --- and $\widetilde{\pi}_i$ is the corresponding conditional law when $f_i(\vect{Z})=1$. We use $\widetilde{\cdot}$ to denote the ``harmonic'' version of a policy. From Definition \ref{def:afeo},
    \begin{align*}
        \afeo(\pi,f,y)&=\abs{\supp_f(\pi)}^{-1}\sum\expectpi{\pi_i}{y_i(\vect{Z})},
    \end{align*}
    where $\pi_i$ is the corresponding conditional law when $f_i(\vect{Z})=1$. Since $y$ is an arbitrary potential outcome function, $\efao(\pi,f,y)=\afeo(\pi,f,y)$ iff (a) $\forall i\in\supp_f(\pi): \expectpi{\widetilde{\pi}}{f_i(\vect{Z})}=\frac{C_\pi}{\abs{\supp_f(\pi)}}$, and (b) $\forall i\in\supp_f(\pi): \expectpi{\widetilde{\pi}_i}{y_i(\vect{Z})}=\expectpi{\pi_i}{y_i(\vect{Z})}$.    
    Condition (a) says that $f$ is homogeneous under $\widetilde{\pi}$. Condition (b) says that $\widetilde{\pi}_i=\pi_i$, which holds generally iff $\pi$ places probability mass on either $\abs{f(\vect{Z})}=0$ or $\abs{f(\vect{Z})}=k$ for some fixed $k>0$, i.e. $f$ is of conditionally fixed cardinality. Consequently, $\widetilde{\pi}$ is equivalent to $\pi$ up to positivity, and by Proposition \ref{thm:cond_homo} $f$ is homogeneous under $\widetilde{\pi}$ iff $f$ is homogeneous under $\pi$.
\end{proof}

\begin{remark}
   The focal map $f(\vect{z})=\vect{z}$ is commuting in the policy space $\mathcal{P}_n$ of positive completely randomized designs i.e. $m$-of-$n$ designs with $m>0$.
\end{remark}

\begin{proposition}\label{thm:commuting_maps_deterministic} A focal map is universally commuting (Definition \ref{def:commuting}) iff it is deterministic (Definition \ref{def:deterministic}).
\end{proposition}

\begin{proof}
    If we show that a deterministic focal map $f$ is the only focal map that satisfies the conditions of Proposition \ref{thm:commuting_fixedcardinality} for any policy $\pi$, then we have shown that it is the only focal map that is universally commuting.
    It can be shown that a deterministic focal map $f$ is homogeneous (Definition \ref{def:homogeneous_focal}) and of conditionally fixed cardinality (Definition \ref{def:fixedcardinality_focal}) under any $\pi$. To show the inverse, we consider a non-deterministic focal map $f$, i.e. $\exists\vect{z}_-,\vect{z}_+\in\booln$ such that $f(\vect{z}_-)\ne f(\vect{z}_+)$. Then we'll construct a policy $\pi$ such that either $\pi$ is not positive w.r.t. $f$ or $f$ is not homogeneous under $\pi$. Let $\pi_{p}(\vect{z})\triangleq p\indicator{\vect{z}=\vect{z}_+} + (1-p)\indicator{\vect{z}=\vect{z}_-}$ be the two-point policy over $\{\vect{z}_-,\vect{z}_+\}$. Let $S_+\triangleq\set{i\in[n]}{f_i(\vect{z}_-)=0,f_i(\vect{z}_+)=1}$, $S_-\triangleq\set{i\in[n]}{f_i(\vect{z}_-)=1,f_i(\vect{z}_+)=0}$, and $S_0\triangleq\set{i\in[n]}{f_i(\vect{z}_-)=1,f_i(\vect{z}_+)=1}$ denote the set of units that respectively gain, lose, and retain their status as a focal node from $\vect{z}_-$ to $\vect{z}_+$. Evidently, $f(\vect{z}_-)\ne f(\vect{z}_+)\implies S_+\cup S_-\ne\emptyset$. That is, at most one of $S_+,S_-$ is empty. Furthermore,
    \begin{align*}
        \forall i\in\supp_f(\pi_p):\expectpi{\pi_p}{f_i(\vect{Z})}=
        \begin{cases}
            p & i\in S_+,\\
            1-p & i\in S_-,\\
            1 & i\in S_0.\\
        \end{cases}
    \end{align*}
    If $S_0\ne\emptyset$, or $S_0=\emptyset,S_+\ne\emptyset,S_-\ne\emptyset$, then for $p\in(0,1)\setminus\{0.5\}$ we get that $\pi_p$ is positive and $f$ is not homogeneous under $\pi_p$. If $S_0=\emptyset,S_+=\emptyset$ then $\pi_1$ is not positive. If $S_0=\emptyset,S_-=\emptyset$ then $\pi_0$ is not positive. This exhausts all possibilities.
\end{proof}

\begin{corollary}
    The all-ones vector $\ones{n}$ is the only focal map which is universally commuting (Definition \ref{def:commuting}) and w.r.t. which every policy $\pi$ is globally positive (Definition \ref{def:positivity}):
    \begin{equation}\label{eq:eao}\tag{$\ast$}
    \forall\mathcal{P}_n,\forall\pi\in\mathcal{P}_n:\efao(\pi,\ones{n},y)=\afeo(\pi,\ones{n},y)\triangleq\eao(\pi,y),
    \end{equation}
    where we refer to $\eao(\pi,y)$ as the expected average outcome under $\pi$.
\end{corollary}

\begin{remark}    
    Unless otherwise stated, in what follows we assume a fixed, arbitrary, potential outcome function $y$ and suppress the dependence of estimands on it.
\end{remark}

\begin{definition}[Policy contrast] \label{def:policy_contrast}
    Consider policies $\pi, \pi'$ that are positive w.r.t. a focal map $f$ (Definition \ref{def:positivity}). Then a policy contrast between $\pi, \pi'$ over $f$ is the difference in expected focal average outcomes (Definition \ref{def:efao}) under both policies: $$\delta_f(\pi, \pi')\triangleq\efao(\pi,f)-\efao(\pi',f).$$
    The policy contrast is said to be trivial if $\pi=\pi'$, and degenerate if $\exists i\in\supp_f(\pi)\cup\supp_f(\pi'): \pi_i\stackrel{f}{\equiv}\pi_i'$ (Definition \ref{def:equivalence_policy}).
\end{definition}

\begin{definition}[Focal contrast] \label{def:focal_contrast}
    Consider policy $\pi$ that is positive w.r.t. focal maps $f, f'$ (Definition \ref{def:positivity}). Then a focal contrast between $f, f'$ under $\pi$ is the difference in average focal expected outcomes (Definition \ref{def:afeo}) over both focal maps: $$\delta_\pi(f, f')\triangleq\afeo(\pi,f)-\afeo(\pi,f').$$
    The focal contrast is said to be trivial if $f=f'$, and degenerate if $\exists i\in\supp_f(\pi)\cup\supp_{f'}(\pi): f_i\stackrel{\pi}{\equiv}f_i'$ (Definition \ref{def:equivalence}).
\end{definition}

\begin{remark}
    Let $\ones{\vect{v}}$ denote the indicator vector for Boolean vector $\vect{v}$. The average treatment effect (ATE) is the focal contrast $\delta_\pi(\ones{\vect{z}=1},\ones{\vect{z}=0})$; when $\pi$ is the all-or-none-treated policy the ATE yields the global average treatment effect (GATE). Consider the interference network to be encoded by the Boolean matrix $\mat{A}$, then $\delta_\pi(\ones{\mat{A}\vect{z}>0},\ones{\mat{A}\vect{z}=0})$ is a common way of defining an average effect of having one or more treated neighbors.
\end{remark}

\begin{definition}[Unit-level contrast] \label{def:unit_contrast}
    A unit-level contrast over $i\in[n]$ is a policy contrast (Definition \ref{def:policy_contrast}) over the $i^\text{th}$ canonical unit vector $\vect{e}_i$; $(\vect{e}_i)_j = 1$ if $j=i$ and $0$ otherwise.
\end{definition}

\begin{definition}[Convex unit-level causal estimand and contrast] \label{def:causal_estimand}
    Let  $w_i\in[0,1]$ such that $ \sum_{i\in [n]} w_i=1$ be known unit-level weights that don't depend on the potential outcome function $y$. An estimand (Definition \ref{def:estimand}), or set of estimands, is said to be a convex unit-level causal estimand in a nontrivial policy space $\mathcal{P}_n$ if $\forall i\in [n]:w_i=0$ or $\exists\,\pi^i_+,\pi^i_-\in\mathcal{P}_n$ with $\pi^i_+\ne\pi^i_-$ such that the estimand yields a convex sum of unit-level contrasts (Definition \ref{def:unit_contrast}): $$\sum_{i\in [n]}{w_i\delta_{\vect{e}_i}(\pi^i_+,\pi^i_-)}=\sum_{i\in [n]}w_i\left(\expectpi{\pi^i_+}{y_i(\vect{Z})}-\expectpi{\pi^i_-}{y_i(\vect{Z})}\right).$$
    The estimand, or set of estimands, is said to be \emph{universally} convex unit-level causal if it is convex unit-level causal in any nontrivial $\mathcal{P}_n$. The corresponding contrast (Definition \ref{def:contrast}) is called a convex unit-level causal contrast.
\end{definition}

\begin{remark}
    We require $\pi^i_+\ne\pi^i_-$ in order for the causal contrast to feel meaningful as an average difference in expected outcomes under two \emph{distinct} policy interventions, so that zero causal contrasts have the appropriate interpretation of the outcome being invariant to the policy change. We appreciate that there can be more or less restrictive ways of defining a unit-level causal estimand --- our definition covers many canonical estimands in the literature like the ATE, conditional ATE, ATE on the treated/untreated, local ATE, etc. --- and causal quantities can go beyond unit-level contrasts.
\end{remark}

\begin{proposition}[Support-aligned non-degenerate focal contrasts are convex unit-level causal]\label{thm:focal_contrast_causal}
    Consider a policy $\pi$ that is positive w.r.t. focal maps $f, f'$ such that $\supp_f(\pi)=\supp_{f'}(\pi)\triangleq S_\pi\supset\emptyset$ (Definition \ref{def:positivity}) and $\diffsupppi{f,f'}(\pi)=S_\pi$ (Definition \ref{def:equivalence}). Then $\exists\mathcal{P}_n'$ such that the focal contrast $\delta_\pi(f,f')$ (Definition \ref{def:focal_contrast}) is convex unit-level causal in $\mathcal{P}_n'$ over $S_{\pi}$ (Definition \ref{def:causal_estimand}).
\end{proposition}

\begin{proof}
    From Definition \ref{def:focal_contrast}, 
    \begin{align*}
        \delta_\pi(f,f')&=\avgpi{i\in \supp_f(\pi)}{\condexpectpi{\pi}{y_i(\vect{Z})}{f_i(\vect{Z})=1}}-\avgpi{i\in \supp_{f'}(\pi)}{\condexpectpi{\pi}{y_i(\vect{Z})}{f'_i(\vect{Z})=1}}\\
        &=\sum_{i\in S_\pi}{\abs{S_\pi}^{-1}\left(\expectpi{\pi_i}{y_i(\vect{Z})}-\expectpi{\pi'_i}{y_i(\vect{Z})}\right)},
    \end{align*}
    where we use the support-alignment of $f,f'$ and $\pi_i,\pi_i'$ are the conditional laws respectively when $f_i(\vect{Z})=1$ and $f'_i(\vect{Z})=1$. Collect all such policies into $\mathcal{P}_n'\triangleq\bigcup_{i\in S_\pi} \{\pi_i\}\cup\{\pi_i'\}$.
    Since $\diffsupp{f}{f'}(\pi)=S_\pi$, $\forall i\in S_{\pi}: \pi_i \ne \pi_i'$. Hence $\mathcal{P}_n'$ 
    contains at least two distinct policies and is nontrivial. Consequently, $\delta_\pi(f,f')$ is convex unit-level causal in $\mathcal{P}_n'$ over $S_\pi$.
\end{proof}

\begin{proposition}\label{thm:focal_contrast_universal_causal}
    Support-aligned non-degenerate focal contrasts (Definition \ref{def:focal_contrast}) are not universally convex unit-level causal (Definition \ref{def:causal_estimand}).
\end{proposition}

\begin{proof}
    We show that one can construct a nontrivial policy space $\mathcal{P}_n$ for which there does not exist a support-aligned non-degenerate focal contrast that is convex unit-level causal in it. Using Definition \ref{def:causal_estimand}, and the notation from the proof for Proposition \ref{thm:focal_contrast_causal}, we note that if the focal contrast $\delta_\pi(f,f')$ is convex unit-level causal in $\mathcal{P}_n$ then $\forall i\in S_\pi: \pi_i=\pi^i_+,$ where $\pi^i_+\in\mathcal{P}_n$ and $\pi_i$ is the conditional law when $f_i(\vect{Z})=1$. Let $\mathcal{P}_n\triangleq\mathcal{P}_n^+$ be the space of positive policies on $\booln$ --- such as the space of i.i.d. treatments with probability $p\in(0,1)$ --- then any policy $\pi^i_+\in\mathcal{P}_n^+$ puts some probability mass on $f_i(\vect{Z})=0$ which would imply that $\nexists\pi$ such that $\forall i:\pi_i=\pi_+^i$, \emph{unless} $f$ is deterministic. By analogous reasoning, the focal maps $f'$ worth considering must also be deterministic. However, two deterministic focal maps $f,f'$ must be identical or not support aligned, i.e. $\nexists \pi,f,f'$ such that the focal contrast $\delta_\pi(f,f')$ is support-aligned, non-degenerate, and convex unit-level causal in $\mathcal{P}_n^+$.
\end{proof}

\begin{definition}[Projective-marginal policy equivalence]\label{def:projmargequiv} Policies $\pi,\pi'$ (Definition \ref{def:policy_space}) are said to be projective-marginally equivalent w.r.t. a focal map $f$ (Definition \ref{def:focal_units}) if: $$\exists c >0,\forall i\in\supp_f(\pi)\cup\supp_f(\pi'):\expectpi{\pi'}{f_i(\vect{Z})}=c\,\expectpi{\pi}{f_i(\vect{Z})}.$$ 
\end{definition}

\begin{proposition}[Support-aligned non-degenerate policy contrasts are convex unit-level causal iff they are w.r.t. focal maps that render harmonic version of policies projective-marginally equivalent] \label{thm:policy_contrast_causal}
    Consider a nontrivial policy space $\mathcal{P}_n$ that is positive w.r.t. focal map $f$ such that $\forall\pi\in\mathcal{P}_n:\supp_f(\pi)=S_f\supset\emptyset$ (Definition \ref{def:positivity}) and $\forall\pi'\in\mathcal{P}_n,\pi'\ne\pi: \diffsupppi{f}(\widetilde{\pi},\widetilde{\pi}') = S_f$ (Definition \ref{def:equivalence_policy}), where $\widetilde{\cdot}$ denotes the harmonic version of a policy. Then the policy contrast $\delta_f(\pi,\pi')$ (Definition \ref{def:policy_contrast}) is convex unit-level causal in $\mathcal{P}_n$ over $S_f$ (Definition \ref{def:causal_estimand}) iff the harmonic version of $\mathcal{P}_n$ is projective-marginally equivalent w.r.t. $f$ (Definition \ref{def:projmargequiv}).
\end{proposition}

\begin{proof}
    As in the proof for Proposition \ref{thm:commuting_fixedcardinality}, definition \ref{def:efao} yields
    \begin{align*}
        \efao(\pi,f,y)&=C_\pi^{-1}\sum_{i\in S_f}\expectpi{\widetilde{\pi}}{f_i(\vect{Z})}\expectpi{\widetilde{\pi}_i}{y_i(\vect{Z})},
    \end{align*}
    where $\widetilde{\pi}$ is the harmonic version of the policy $\pi$, whose probability law is given by: $$\probpi{\widetilde{\pi}}{\cdot}\triangleq C_\pi \sum_{k=1}^\infty k^{-1}\condprobpi{\pi}{\cdot}{\abs{f(\vect{Z})}=k}\condprobpi{\pi}{\abs{f(\vect{Z})}=k}{f(\vect{Z})\ne\vect{0}},$$
    where $C_\pi\triangleq\left(\sum_{k=1}^\infty k^{-1}\condprobpi{\pi}{\abs{f(\vect{Z})}=k}{f(\vect{Z})\ne\vect{0}}\right)^{-1}$ is the conditional harmonic mean of the number of focal nodes --- conditional on them being positive --- and $\widetilde{\pi}_i$ is the corresponding conditional law when $f_i(\vect{Z})=1$. Definition \ref{def:policy_contrast} --- and the support alignment of $\pi,\pi'$ --- yield the policy contrast $$\delta_f(\pi,\pi')=\sum_{i\in S_f}C_\pi^{-1}\expectpi{\widetilde{\pi}}{f_i(\vect{Z})}\expectpi{\widetilde{\pi}_i}{y_i(\vect{Z})} - C_{\pi'}^{-1}\expectpi{\widetilde{\pi}'}{f_i(\vect{Z})}\expectpi{\widetilde{\pi}'_i}{y_i(\vect{Z})},$$ which can be written as a strictly convex sum over units in $S_f$ iff \sloppy (a) $C_\pi^{-1}\sum\expectpi{\widetilde{\pi}}{f_i(\vect{Z})} = C_{\pi'}^{-1}\sum\expectpi{\widetilde{\pi}'}{f_i(\vect{Z})}=1$, and (b) $\forall i\in S_f: \frac{\expectpi{\widetilde{\pi}}{f_i(\vect{Z})}}{\expectpi{\widetilde{\pi}'}{f_i(\vect{Z})}}=\frac{C_\pi}{C_{\pi'}}$. Condition (a) already holds since:$$C_\pi^{-1}\sum\expectpi{\widetilde{\pi}}{f_i(\vect{Z})}=C_\pi^{-1}\expectpi{\widetilde{\pi}}{\abs{f(\vect{Z})}}=1,$$ and similarly for $\pi'$. Condition (b) further holds iff $\widetilde{\pi},\widetilde{\pi}'$ are projective-marginally equivalent w.r.t. $f$. Finally, since $\diffsupppi{f}(\widetilde{\pi},\widetilde{\pi}')=S_f$, $\forall i\in S_f:\widetilde{\pi}_i\ne\widetilde{\pi}'_i$. Consequently, $\delta_f(\pi,\pi')$ is convex unit-level causal in $\mathcal{P}_n$ over $S_f$.
\end{proof}

\begin{remark}
    Policy contrasts w.r.t. the focal map $f(\vect{z})=\vect{z}$ are unit-level causal under completely randomized or $m$-of-$n$ designs with $m>0$.
\end{remark}

\begin{remark}
    Propositions \ref{thm:focal_contrast_causal} and \ref{thm:policy_contrast_causal} show that while focal contrasts are generically convex unit-level causal, policy contrasts need stronger assumptions on the focal map and the policy space to be convex unit-level causal.
\end{remark}

\begin{proposition}\label{thm:policy_contrast_det_focal_causal}
    Nontrivial policy contrasts (Definition \ref{def:policy_contrast}) are universally convex unit-level causal (Definition \ref{def:causal_estimand}) iff they are taken over deterministic focal maps (Definition \ref{def:deterministic}).
\end{proposition}

\begin{proof}
    If we show that a deterministic focal map $f$ is the only focal map that satisfies the conditions of Proposition \ref{thm:policy_contrast_causal} for any policy pair $\pi,\pi'$, then we have shown that it is the only focal map that renders policy contrasts as universally convex unit-level causal.
    It can be shown that a deterministic focal map $f$ renders any $\pi,\pi'$ as projective-marginally equivalent (Definition \ref{def:projmargequiv}). To show the inverse, we consider a non-deterministic focal map $f$, i.e. $\exists\vect{z}_-,\vect{z}_+\in\booln$ such that $f(\vect{z}_-)\ne f(\vect{z}_+)$. Then we'll construct policies $\pi\ne\pi'$ such that either $\pi$ or $\pi'$ is not positive w.r.t. $f$, or $\pi,\pi'$ are harmonic versions of some policies but are not projective-marginally equivalent w.r.t. $f$. Similarly to the proof for Proposition \ref{thm:commuting_maps_deterministic}, let $\pi_{p}(\vect{z})\triangleq p\indicator{\vect{z}=\vect{z}_+} + (1-p)\indicator{\vect{z}=\vect{z}_-}$ be the two-point policy over $\{\vect{z}_-,\vect{z}_+\}$. It can be shown that $\pi_p=\widetilde{\pi}_{h(p)}$ where $h(p)\triangleq \left(1+\left(\frac{1}{p}-1\right)\frac{\abs{f(\vect{z}_-)}}{\abs{f(\vect{z}_+)}}\right)^{-1}$. That is, $\pi_p$ is a harmonic version of another two-point policy $\pi_{h(p)}$, and therefore sufficient to consider for the purposes of this proof.
    
    Let $S_+\triangleq\set{i\in[n]}{f_i(\vect{z}_-)=0,f_i(\vect{z}_+)=1}$, $S_-\triangleq\set{i\in[n]}{f_i(\vect{z}_-)=1,f_i(\vect{z}_+)=0}$, and $S_0\triangleq\set{i\in[n]}{f_i(\vect{z}_-)=1,f_i(\vect{z}_+)=1}$ denote the set of units that respectively gain, lose, and retain their status as a focal node from $\vect{z}_-$ to $\vect{z}_+$. Evidently, $f(\vect{z}_-)\ne f(\vect{z}_+)\implies S_+\cup S_-\ne\emptyset$. That is, at most one of $S_+,S_-$ is empty. Furthermore,
    \begin{align*}
        \forall i\in\supp_f(\pi_p):\expectpi{\pi_p}{f_i(\vect{Z})}=
        \begin{cases}
            p & i\in S_+,\\
            1-p & i\in S_-,\\
            1 & i\in S_0.\\
        \end{cases}
    \end{align*}
    If $S_0\ne\emptyset$, or $S_0=\emptyset,S_+\ne\emptyset,S_-\ne\emptyset$, then for $p\ne q$ we get that $\{\pi_p,\pi_q\}$ is positive but not projective-marginally equivalent w.r.t. $f$, i.e. $f$ is not convex unit-level causal in $\{\pi_p,\pi_q\}$ . If $S_0=\emptyset,S_+=\emptyset$ then $\pi_1$ is not positive and if $S_0=\emptyset,S_-=\emptyset$ then $\pi_0$ is not positive, i.e. if $S_0\ne\emptyset$ then $\{\pi_0,\pi_1\}$ is not positive w.r.t. $f$. This exhausts all possibilities.
\end{proof}

\begin{remark}
    Propositions \ref{thm:commuting_maps_deterministic} and \ref{thm:policy_contrast_det_focal_causal} show that deterministic focal maps are unique in that they are universally commuting \emph{and} their nontrivial policy contrasts are universally convex unit-level causal.
\end{remark}

\begin{definition}[Expected welfare function]\label{def:welfare} An expected welfare function is an estimand (Definition \ref{def:estimand}), i.e. a functional on the joint law of outcomes and focal units, that captures a target state of the world. \end{definition}

\begin{definition}[Estimand (set) sufficient for policy choice] \label{def:policy_relevant}
    A set of estimands $T=\{\theta_1,\theta_2, \dots,\theta_{\abs{T}}\}$ (Definition \ref{def:estimand}) is said to be sufficient for policy choice in the nontrivial policy space $\mathcal{P}_n$ w.r.t. an expected welfare function $W:\mathcal{P}_n\times\mathcal{F}_n\times\mathcal{Y}_n\to\real$ (Definition \ref{def:welfare}) if there exists a known function $h:\mathcal{P}_n\times\real^{\abs{T}}\to\real$ that yields a policy that maximizes the expected welfare: $$\argmax_{\pi\in\mathcal{P}_n}h(\pi,\theta_1(\pi,f_1,y),\theta_2(\pi,f_2,y),\dots,\theta_{\abs{T}}(\pi,f_{\abs{T}},y))\subseteq\argmax_{\pi\in\mathcal{P}_n}W(\pi,f,y).$$
    It is said to be \emph{universally} sufficient for policy choice w.r.t. $W$ if it is sufficient for any nonempty $\mathcal{P}_n$ w.r.t. $W$.
\end{definition}

\begin{definition}[Identification by estimand (set)]\label{def:identification}
    A set of estimands $T=\{\theta_1,\theta_2, \dots,\theta_{\abs{T}}\}$ (Definition \ref{def:estimand}) is said to identify an estimand $\theta:\mathcal{P}_n\times\mathcal{F}_n\times\mathcal{Y}_n\to\real$ in the nonempty policy space $\mathcal{P}_n$ w.r.t. the potential outcome space $\mathcal{Y}_n$ such that $\abs{\mathcal{Y}_n}\ge 2$ if $\forall y,y'\in\mathcal{Y}_n$ and $\forall \pi\in\mathcal{P}_n:$
    $$\{\theta_1(\pi,f_1,y),\dots,\theta_{\abs{T}}(\pi,f_{\abs{T}},y)\}=\{\theta_1(\pi,f_1,y'),\dots,\theta_{\abs{T}}(\pi,f_{\abs{T}},y')\}\implies\theta(\pi,f,y)=\theta(\pi,f,y').$$
    $T$ is said to \emph{functionally} identify $\theta$ in $\mathcal{P}_n$ w.r.t. $\mathcal{Y}_n$ if there exists a \emph{known} function $\phi:\mathcal{P}_n\times\real^{\abs{T}}\to\real:$ $$\forall y\in\mathcal{Y}_n,\forall\pi\in\mathcal{P}_n:\phi(\pi,\theta_1(\pi,f_1,y),\theta_2(\pi,f_2,y),\dots,\theta_{\abs{T}}(\pi,f_{\abs{T}},y))=\theta(\pi,f,y).$$
\end{definition}

\begin{remark}
    When we refer to identification without referencing a policy space $\mathcal{P}_n$ and/or potential outcome space $\mathcal{Y}_n$ then it is with reference to an arbitrary policy and/or potential outcome space.
\end{remark}

\begin{proposition}\label{thm:constant_pairwise_id}
    An estimand set $T$ (functionally) identifies an estimand $\theta$ in $\mathcal{P}_n$ up to a constant iff it (functionally) identifies pairwise differences in $\theta$ in $\mathcal{P}_n$.
\end{proposition}

\begin{proof}
    Let $T(\pi,y)\triangleq\{\theta_1(\pi,f_1,y),\theta_2(\pi,f_2,y),\dots,\theta_{\abs{T}}(\pi,f_{\abs{T}},y)\}$ be the estimand set. For the inverse direction, say $T$ identifies $\theta$ up to a constant i.e. $\forall\pi\in\mathcal{P}_n$:
    $$T(\pi,y)=T(\pi,y')\implies\theta(\pi,f,y)=\theta(\pi,f,y')+c_{y,y'}.$$
    Considering all pairs of policies $\forall\pi,\pi'\in\mathcal{P}_n$ we immediately get:
    $$T(\pi,y)=T(\pi,y'),T(\pi',y)=T(\pi',y')\implies\theta(\pi,f,y)-\theta(\pi',f,y)=\theta(\pi,f,y')-\theta(\pi',f,y')=c_{y,y'}.$$    
    For the forward direction, say $T$ identifies pairwise differences in $\theta$. Then there must exist a pair of policies $\pi,\pi'\in\mathcal{P}_n$ for which the antecedant of the previous implication holds --- else the implication is vacuously true. This yields the first implication for both $\pi$ and $\pi'$. For functional identification, first consider the inverse direction: there exists a known function $\phi: \theta(\pi,f,y)=\phi(T(\pi,y))+c_y\implies \theta(\pi,f,y)-\theta(\pi',f,y)=\phi(T(\pi,y))-\phi(T(\pi',y))=\phi'(T(\pi,y),T(\pi',y))$ where $\phi'(u,v)\triangleq\phi(u)-\phi(v)$ is known. For the forward direction, assume there exists a known function $\phi'$ such that $\theta(\pi,f,y)-\theta(\pi',f,y)=\phi(T(\pi,y),T(\pi',y))$. Fix an arbitrary reference policy $\pi_0\in\mathcal{P}_n$, then there exists a known function $\phi:\theta(\pi,f,y)=\phi(T(\pi,y),T(\pi_0,y))+\theta(\pi_0,f,y)=\phi'(T(\pi,y))+c_y$, where $\phi'(u)=\phi(u,T(\pi_0,y))$ is known.
\end{proof}

\begin{proposition}\label{thm:universal_policychoice} An estimand set $T$ is universally sufficient for policy choice (Definition \ref{def:policy_relevant}) w.r.t. expected welfare function $W$ (Definition \ref{def:welfare}) iff $W$ is a weakly monotone transformation of an estimand functionally identified by $T$ (Definition \ref{def:identification}).
\end{proposition}

\begin{proof}
    Let $T(\pi,y)\triangleq\{\theta_1(\pi,f_1,y),\theta_2(\pi,f_2,y),\dots,\theta_{\abs{T}}(\pi,f_{\abs{T}},y)\}$ be the estimand set. For the forward direction, say $W$ is a weakly monotone transformation of an estimand functionally identified by $T$, then there exist functions $\psi, \phi$ such that $\psi$ is weakly monotone, $\phi$ is known, and $W(\pi, f, y)=\psi(\phi(T(\pi, y)))$: 
    \begin{align*}
    \argmax W(\pi,f,y) = \argmax\psi(\phi(T(\pi, y))) \supseteq
    \begin{cases}
        \argmax\phi(T(\pi, y)) & \psi\text{ non-decreasing},\\
        \argmax-\phi(T(\pi, y)) & \psi\text{ non-increasing}.\\
    \end{cases}
    \end{align*}
    For the inverse direction, say $T$ is universally sufficient for policy choice, then there exists a known function $h$ such that: $\argmax_{\pi\in\mathcal{P}_n}h\left(\pi,T(\pi,y)\right)\subseteq\argmax_{\pi\in\mathcal{P}_n}W(\pi,f,y).$ Say $\mathcal{P}_n=\{\pi,\pi'\}$ then 
    \begin{align*}
        h\left(\pi,T(\pi,y)\right)> h\left(\pi',T(\pi',y)\right)\implies W(\pi,f,y)\ge W(\pi',f,y),\\
        h\left(\pi,T(\pi,y)\right)= h\left(\pi',T(\pi',y)\right)\implies W(\pi,f,y)= W(\pi',f,y).
    \end{align*}
    Since $T$ is universally sufficient for policy choice, this holds for any policy pair. Therefore, there exists some non-decreasing function $\psi$ such that $W(\pi,f,y)=\psi(h(\pi,T(\pi,y)))$, with $h$ known. 
\end{proof}

\begin{definition}\label{def:linear_welfare}
    From Definition \ref{def:efao} for EFAO, utilitarian welfare is encoded by the expected welfare function:
    $$W_\textsf{U}^C(\pi,f,y)\triangleq\efao(\pi,f,y)-C(\pi),$$
    where $C:\mathcal{P}_n\to\real\cup\{-\infty,\infty\}$ is a known function encoding the cost of deploying a policy in $\mathcal{P}_n$. $W_\textsf{U}^0$ is said to be \emph{costless} utilitarian welfare. $W_\textsf{U}^C(\pi,\ones{n},y)$ is said to be \emph{total} utilitarian welfare.
\end{definition}

\begin{remark}
    Can be generalized to weighted utilitarian welfare by appropriately rescaling unit-level outcomes.
\end{remark}

\begin{corollary}\label{thm:efao_policychoice_costless}
    An estimand set $T$ is universally sufficient for policy choice (Definition \ref{def:policy_relevant}) w.r.t. costless utilitarian welfare $W_\textsf{U}^0$ (Definition \ref{def:linear_welfare}) iff the EFAO is a weakly monotone transformation of an estimand functionally identified by $T$ (Definition \ref{def:identification}).
\end{corollary}

\begin{proof}
    Follows by applying Proposition \ref{thm:universal_policychoice} to $W_\textsf{U}^0$.
\end{proof}

\begin{proposition}\label{thm:efao_policychoice}
    An estimand set $T$ is universally sufficient for policy choice (Definition \ref{def:policy_relevant}) w.r.t. utilitarian welfare with arbitrary costs $W_\textsf{U}^C$ (Definition \ref{def:linear_welfare}) iff the EFAO is functionally identified by $T$ up to a constant (Definition \ref{def:identification}).
\end{proposition}

\begin{proof}
    For the forward direction, note that since $C$ is a known function and $\efao(\pi,f,y)$ is functionally identified by $T$ up to a constant, $W_\textsf{U}^C(\pi,f,y)\triangleq \efao(\pi,f,y)-C(\pi)$ is a constant shift from an estimand functionally identified by $T$, and Proposition \ref{thm:universal_policychoice} yields the result. For the inverse direction, say $T$ is universally sufficient for policy choice w.r.t. $W_\textsf{U}^C$. For the sake of contradiction, assume EFAO is \emph{not} identified by $T$ up to a constant. From Proposition \ref{thm:constant_pairwise_id}, $\exists y,y',\exists\pi,\pi':T(\pi,y)=T(\pi,y'), T(\pi',y)=T(\pi',y')$ and $\efao(\pi,f,y)-\efao(\pi,f,y')\ne \efao(\pi',f,y)-\efao(\pi',f,y')$. Without loss of generality, assume $\efao(\pi,f,y)-\efao(\pi,f,y') < \efao(\pi',f,y)-\efao(\pi',f,y')$ and pick a cost function such that: $$C(\pi)-C(\pi')\in\left(\efao(\pi,f,y)-\efao(\pi',f,y), \efao(\pi,f,y')-\efao(\pi',f,y')\right),$$ which is possible since $C$ is arbitrary. For the policy space $\mathcal{P}_n\triangleq\{\pi,\pi'\}$:
    \begin{align*}
        \argmax_{\pi\in\mathcal{P}_n} W_\textsf{U}^C(\pi,f,y)=\pi'\ne\argmax_{\pi\in\mathcal{P}_n} W_\textsf{U}^C(\pi,f,y')=\pi.
    \end{align*}
    Since $T(\pi,y)=T(\pi,y'), T(\pi',y)=T(\pi',y')$, any function $h(\pi,T(\pi,y))$ will pick the same policy for $y$ and $y'$, i.e. $T$ is \emph{not} sufficient for policy choice in $\mathcal{P}_n$ w.r.t. $W_\textsf{U}^C$ which contradicts our assumption of $T$ being universally sufficient for policy choice w.r.t. $W_\textsf{U}^C$. Consequently, EFAO is identified by $T$ up to a constant. It remains to be shown that the corresponding function is known. Say the identification is \emph{not} functional, and the difference $\efao(\pi,f,y)-\efao(\pi',f,y)$ cannot be written as a known function of $T(\pi,y),T(\pi',y)$. Then, by picking a cost function that flips the policy choice implied by any known function of $T$ --- as in the argument above -- we can show that there cannot exist a known function $h$ for all cost functions $C$ such that $\argmax h(\pi,T(\pi,y))\subseteq\argmax W_\textsf{U}^C(\pi,f,y)$ implying that $T$ is not universally sufficient for policy choice w.r.t. $W_\textsf{U}^C$. Therefore, EFAO is functionally identified by $T$ up to a constant.
\end{proof}

\begin{definition}[Exposure map] \label{def:exposure_map}
    An exposure map is a function $d:\booln\to\mathcal{D}^n$ that maps a treatment vector to an exposure vector with values in the exposure space $\mathcal{D}$. $d$ is said to be a \emph{discrete} exposure map if the exposure space $\mathcal{D}$ is discrete (and usually of low cardinality  $\abs{\mathcal{D}}\ll n$).
\end{definition}

\begin{remark} A focal map (Definition \ref{def:focal_units}) is an exposure map (Definition \ref{def:exposure_map}) with codomain $\booln$.
\end{remark}

\begin{remark} This definition of an exposure map matches that in \citet{savje2023exposure}. \citet{aronow2017interference} instead define an exposure map as a common (i.e. not unit-dependent) function that takes units' traits as an argument.
\end{remark}

\begin{definition}[Focal space of exposure map]\label{def:focal_exposure}
    For exposure map $d:\booln\to\mathcal{D}^n$ (Definition \ref{def:exposure_map}), $\set{\ones{d(\vect{z})=e}}{\forall e\in\mathcal{D}}$ is said to its focal space.
\end{definition}

\begin{definition}[Positive policy w.r.t. exposure map]\label{def:positivity_exposure} 
    Let $\mathbb{F}_d$ be the focal space defined by the exposure map $d$ (Definition \ref{def:focal_exposure}). A policy $\pi$ is said to be (globally) positive w.r.t. $d$ if it is (globally) positive w.r.t. $\mathbb{F}_d$ (Definition \ref{def:positivity}). 
\end{definition}

\begin{definition}[Average outcomes by exposure (AOE)]\label{def:avg_exposure_outcome}
    The average outcomes by exposure (AOE) over exposure map $d$ under policy $\pi$, when $\pi$ is positive w.r.t. $d$ (Definition \ref{def:positivity_exposure}), is given by the set of $\abs{\mathcal{D}}$ average focal expected outcomes (AFEO; Definition \ref{def:afeo}) over the focal space of $d$ (Definition \ref{def:focal_exposure}) under $\pi$: $$\aoe(\pi, d)\triangleq\set{\afeo(\pi,\ones{d(\vect{z})=e})}{\forall e\in\mathcal{D}, \pi\text{ positive w.r.t. }\ones{d(\vect{z})=e}}.$$
\end{definition}

\begin{proposition}[EAO is identified by AOE iff unit-homogeneous exposure distributions]\label{thm:eao_identification}
    $\eao(\pi)$ in Eq. \eqref{eq:eao} is identified (Definition \ref{def:identification}) by the AOE over exposure map $d$ (Definition \ref{def:avg_exposure_outcome}) under policy $\pi$ iff the exposure distribution $\probpi{\pi}{d_i(\vect{Z})}$ is unit-homogeneous i.e. independent of $i$.
\end{proposition}

\begin{proof}
    \sloppy For the forward direction, assume that the exposure distribution under $\pi$ is homogeneous and encoded by the probability measure $\mu^\pi$, then from Eq. \eqref{eq:eao} and Definition \ref{def:avg_exposure_outcome}:
    \begin{align*}        
        \eao(\pi)&=n^{-1}\sum_{i\in [n]}\expectpi{\pi}{y_i(\vect{Z})}=n^{-1}\sum_{i\in [n]}\int_{\mathcal{D}}\condexpectpi{\pi}{y_i(\vect{Z})}{d_i(\vect{Z})=e}\,d\mu^\pi(e)\\
        &=\int_{\mathcal{D}}\avgpi{i\in[n]}{\condexpectpi{\pi}{y_i(\vect{Z})}{d_i(\vect{Z})=e}}\, d\mu^\pi(e) = h(\pi,\aoe(\pi, d)),
    \end{align*}
    where $h(\pi,\cdot)$ is the expectation of a measurable function of exposures under $\pi$ i.e. the EAO is identified by the AOE. To show the inverse, we will explicate the dependence of estimands on the potential outcome function $y$ and show that under a heterogeneous exposure distribution, encoded by the probability measures $\mu^\pi_i$, $\exists y,y'$ such that the $\aoe(\pi,d,y)=\aoe(\pi,d,y')$ but $\eao(\pi,y)\ne\eao(\pi,y')$ i.e. the EAO is \emph{not} identified by the AOE. Since exposure distributions are heterogeneous $\exists i,j\in[n],\exists D\subset\mathcal{D}:\mu^\pi_i(D)\ne\mu^\pi_j(D)$. Let $y_i(\vect{z})=\indicator{d_i(\vect{z})\in D}$ and $y'_i(\vect{z})=c_i\indicator{d_i(\vect{z})\in D}$ for some $c_i\in\real$.
    \begin{align*}
        \aoe(\pi,d,y)&=\set{\indicator{e\in D}}{e\in\mathcal{D}},\\
        \aoe(\pi,d,y')&=\set{\avgpi{i\in[n]}{c_i}\indicator{e\in D}}{e\in\mathcal{D}},\\
        \eao(\pi,y)&=n^{-1}\sum_{i\in [n]}\int_{\mathcal{D}}\condexpectpi{\pi}{y_i(\vect{Z})}{d_i(\vect{Z})=e}\,d\mu^\pi_i(e) = \avgpi{i\in[n]}{\mu^\pi_i(D)},\\
        \eao(\pi,y')&= \avgpi{i\in[n]}{c_i\mu^\pi_i(D)}.
    \end{align*}
    With $\delta>0$ set $c_i=1+\delta, c_j=1-\delta, c_k=1$ for $k\ne i,j$, implying $\aoe(\pi,d,y)=\aoe(\pi,d,y')$ and $\eao(\pi,y')-\eao(\pi,y)=\frac{\delta}{n}(\mu^\pi_i(D)-\mu^\pi_j(D))\ne 0$.
\end{proof}

\begin{definition}[Correctly specified exposure map] \label{def:correct_exposure}
    An exposure map $d$ (Definition \ref{def:exposure_map}) is said \citep[after][Condition 1]{aronow2017interference} to be correctly specified if $\forall\vect{z},\vect{z}'\in\booln,\forall i\in[n]:d_i(\vect{z})=d_i(\vect{z}')\implies y_i(\vect{z})=y_i(\vect{z}')$. Else $d$ is said to be misspecified.    
\end{definition}

\begin{remark}
    Being correctly specified does not render an exposure map automatically meaningful. For instance, $d\triangleq y$ is a trivially correctly specified exposure map, but it correlates perfectly with the outcomes. Likewise, given a correctly specified exposure map, unit-specific garblings of the map remain correctly specified. For example, if $d_i$ gives the number of treated neighbors of $i$ and is correctly specified, then some arbitrary relabeling, e.g., $d^*_i(\vect{z}) = r_i(d_i(\vect{z}))$, where $r_i$ is any bijective function, is also correctly specified.
\end{remark}

\begin{remark}
    Propositions \ref{thm:efao_policychoice} and \ref{thm:eao_identification} show that AOEs are \emph{not} universally sufficient for policy choice w.r.t. total utilitarian welfare with arbitrary costs, but they are sufficient in policy spaces where the exposure distribution is unit-homogeneous. For instance, AOE over the (correct) exposure map of ``number of treated neighbors'' is sufficient for policy choice in the space of Bernoulli designs for regular graphs. Similarly, AOE over the (correct) exposure map of ``number of up to $k$ treated neighbors'', where $k$ is the minimum degree of the graph, is sufficient for policy choice in the space of all-or-none-treated designs for any graph.
\end{remark}

\begin{proposition}\label{thm:correctly_specified_exposure}
    Consider a correctly specified exposure map $d$ (Definition \ref{def:correct_exposure}), its focal space $\mathbb{F}_d$ (Definition \ref{def:focal_exposure}), and a policy space $\mathcal{P}_n$. From Definitions \ref{def:avg_exposure_outcome} and \ref{def:positivity}: $$\forall \pi,\pi'\in\mathcal{P}_n: \supp_{f}(\pi) = \supp_{f}(\pi')\,\forall f\in\mathbb{F}_d\implies \aoe(\pi,d)=\aoe(\pi',d).$$
\end{proposition}

\begin{proof}
    The proof follows straightforwardly from the definitions.
\end{proof}

\begin{corollary}[Focal contrasts from a correctly specified exposure map are convex unit-level causal in the space of deterministic policies]\label{thm:focal_contrast_correct_exposure}
    Consider a support-aligned non-degenerate focal contrast $\delta_\pi(f,f')$ satisfying the conditions of Proposition \ref{thm:focal_contrast_causal}, where $f=\ones{d(\vect{z})=e}$ and $f'=\ones{d(\vect{z})=e'}$ for distinct exposures $e\ne e'$ of a correctly specified exposure map $d$ (Definition \ref{def:correct_exposure}). Then the policy space $\mathcal{P}_n'$ constructed in Proposition \ref{thm:focal_contrast_causal} can be taken to consist of deterministic interventions. Thus $\delta_\pi(f,f')$ is convex unit-level causal (Definition \ref{def:causal_estimand}) in any policy space that includes all deterministic interventions.
\end{corollary}

\begin{proof}
    Since $d$ is correctly specified, $y_i(\vect{z})$ is constant over $\set{\vect{z}}{d_i(\vect{z})=e}$; denote this value $y_i(e)$. By Proposition \ref{thm:correctly_specified_exposure}, the conditional expectations in the focal contrast collapse: $\condexpectpi{\pi}{y_i(\vect{Z})}{f_i(\vect{Z})=1}=y_i(e)$ and $\condexpectpi{\pi}{y_i(\vect{Z})}{f'_i(\vect{Z})=1}=y_i(e')$. Let $\supp_f(\pi)=\supp_{f'}(\pi)\triangleq S_\pi$. For each $i\in S_\pi$, since $i$ has positive probability of achieving both exposures, there exist $\vect{z}^i_+,\vect{z}^i_-\in\booln$ with $d_i(\vect{z}^i_+)=e$ and $d_i(\vect{z}^i_-)=e'$. The deterministic policies assigning $\vect{z}^i_+$ and $\vect{z}^i_-$ are distinct (since $e\ne e'$) and yield expected outcomes $y_i(e)$ and $y_i(e')$ respectively for unit $i$, so they witness that the contrast is convex unit-level causal.
\end{proof}

\begin{figure}
    \centering
    \includegraphics[width=0.75\textwidth]{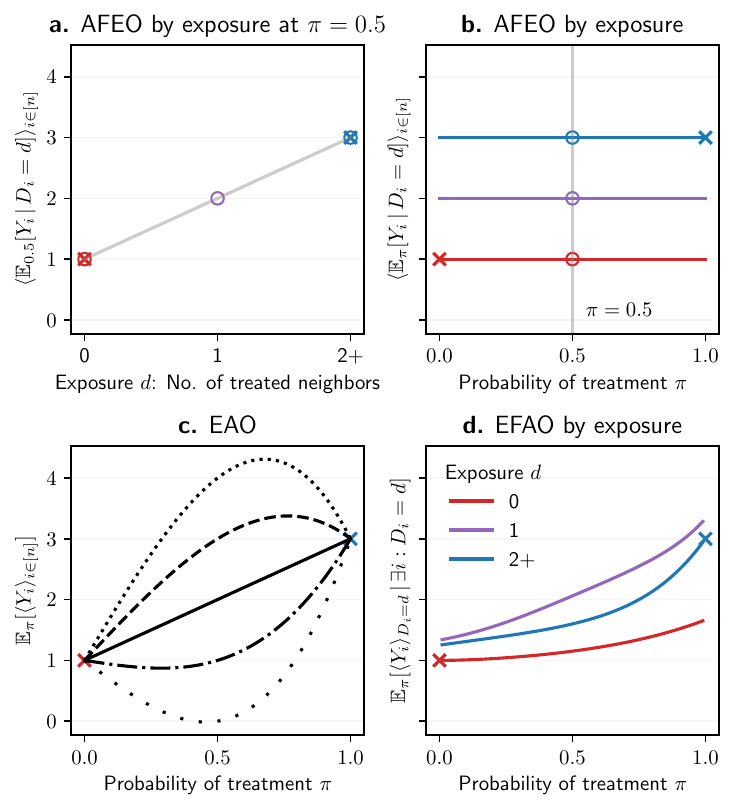}
    \caption{For a correctly specified exposure map, panel (b) shows that the AOE (Definition \ref{def:avg_exposure_outcome}), i.e. AFEO by exposure, is not a function of the treatment policy $\pi$, as suggested by Proposition \ref{thm:correctly_specified_exposure}. See the caption of Figure \ref{fig:aoe_vs_eao} for further details on these dose--response curves.}
    \label{fig:dose_response_curves_correctexposures}
\end{figure}

\setcounter{proposition}{11}
\begin{corollary}\label{thm:correctly_specified_offpolicy}
    Consider a correctly specified exposure map $d$ and a policy space $\mathcal{P}_n$ that is globally positive w.r.t. $d$ (Definition \ref{def:positivity_exposure}). Then $\forall\pi,\pi'\in\mathcal{P}_n$, $\eao(\pi')$ in Eq. \eqref{eq:eao} is identified (Definition \ref{def:identification}) by $\aoe(\pi,d)$ (Definition \ref{def:avg_exposure_outcome}) iff the exposure distribution $\probpi{\pi'}{d_i(\vect{Z})}$ is unit-homogeneous, and: $$\eao(\pi')=h(\pi',\aoe(\pi,d)),$$
    where $h(\pi',\cdot)$ is the expectation of a measurable function of exposures under $\pi'$.
\end{corollary}
\setcounter{proposition}{12}

\begin{proof}
    The proof follows by combining the results in Proposition \ref{thm:eao_identification} and Proposition \ref{thm:correctly_specified_exposure}.
\end{proof}

\begin{remark}
    Corollary \ref{thm:correctly_specified_offpolicy} allows for identification of the EAO at \emph{any point} in a policy space that is globally positive w.r.t. a correctly specified exposure map $d:\booln\to\mathcal{D}^n$, using the AOE over $d$ at a \emph{single} point in that policy space, but only under the special condition of unit-homogeneous exposure distributions. If the exposure distributions are \emph{not} unit-homogeneous, then one can solve a linear program for the unknown functionals $\mathsf{y}_i(e)\in\mathscr{Y}$ under the constraint that: $$\set{\avgpi{i\in[n]}{\mathsf{y}_i(e)}}{\forall e\in\mathcal{D}}=\aoe(\pi,d),$$
    to obtain bounds for the EAO:
    $$\eao(\pi')\in\left[\min_{\mathsf{y}_i(e)\in\mathscr{Y}}\avgpi{i\in[n]}{\expectpi{\pi'}{\mathsf{y}_i(e)}},\max_{\mathsf{y}_i(e)\in\mathscr{Y}}\avgpi{i\in[n]}{\expectpi{\pi'}{\mathsf{y}_i(e)}}\right].$$
    See, for example, Figure \ref{fig:cai_partialidentification}. By considering restricted spaces $\mathscr{Y}$ --- say by assuming the unit-level outcomes $\mathsf{y}_i(e)$ are monotonic in exposure $e$ --- one can potentially obtain tighter bounds, but if they are too constrained then there may not exist any feasible solutions.
\end{remark}

\begin{definition}[Degenerate exposure (map)]\label{def:degenerate} 
    An exposure map $d:\booln\to\mathcal{D}^n$ is said to be degenerate if $\exists\pi$ such that $\pi$ is positive w.r.t. $d$ but $\nexists\pi$ such that $\pi$ is globally positive w.r.t. $d$ (Definition \ref{def:positivity_exposure}). The corresponding exposures $e\in\mathcal{D}$ for which $\exists\pi$ such that $\pi$ is positive w.r.t. the focal map $\ones{d(\vect{z})=e}$ but $\nexists\pi$ such that $\pi$ is globally positive w.r.t. $\ones{d(\vect{z})=e}$ are called degenerate exposures.
\end{definition}

\begin{remark}
    The exposure map of ``number of treated neighbors'' in a non-regular graph is degenerate. The exposure of ``$k$ treated neighbors'' under that exposure map is degenerate in a graph containing nodes with degree less than $k$.
\end{remark}

\setcounter{proposition}{11}
\begin{corollary}\label{thm:eao_degenerate}
    $\eao(\pi)$ is not identified by AOEs over a degenerate exposure map $d$ (Definition \ref{def:degenerate}) if $\pi$ places a positive probability over degenerate exposures.
\end{corollary}
\setcounter{proposition}{12}

\begin{proof}
    The proof follows by noting that the exposure probability $\probpi{\pi}{d_i(\vect{Z})=e}$ for a degenerate exposure $e$ is zero for some but not all units i.e. the exposure distribution is not unit-homogeneous.
\end{proof}

\begin{definition}[Coarsened exposure map]\label{def:coarsened}
    An exposure map $d$ (Definition \ref{def:exposure_map}) is said to be a coarsening of another map $\widetilde{d}$ if $\forall\vect{z},\vect{z}'\in\booln,\forall i\in[n]:\widetilde{d}_i(\vect{z})=\widetilde{d}_i(\vect{z}')\implies d_i(\vect{z})=d_i(\vect{z}')$.
\end{definition}

\begin{definition}[Over-specified exposure map]\label{def:overspecified}
    An exposure map $d$ is said to be over-specified if it is correctly specified (Definition \ref{def:correct_exposure}), degenerate (Definition \ref{def:degenerate}), and can be coarsened (Definition \ref{def:coarsened}) into a correctly specified map that is not degenerate.
\end{definition}

\begin{remark}
    The exposure map of ``number of treated neighbors'' in a non-regular graph is over-specified if the true exposure map is the ``number of up to $k$ treated neighbors,'' where $k$ is the minimum degree in the graph.
\end{remark}

\begin{remark} Definition \ref{def:overspecified} and Corollary \ref{thm:eao_degenerate} suggest that it is possible to \emph{over-specify} an exposure map to an extent that renders EAO non-identifiable and therefore AOEs not universally sufficient for policy choice.
    
\end{remark}

\end{document}

%% file: commands.tex
\newcommand{\bool}[0]{\{0,1\}}
\newcommand{\booln}[0]{\bool^n}
\newcommand{\real}[0]{\mathbb{R}}

\newcommand{\vect}[1]{\boldsymbol{#1}}
\newcommand{\ones}[1]{\vect{1}_{#1}}

\newcommand{\mat}[1]{\mathbf{#1}}
\newcommand{\abs}[1]{\left\vert#1\right\rvert}
\newcommand{\set}[2]{\left\{#1 \, \middle\vert \, #2\right\}}
\newcommand{\indicator}[1]{\left[#1\right]}

\newcommand{\avg}[1]{\left\langle#1\right\rangle}
\newcommand{\avgpi}[2]{\left\langle#2\right\rangle_{#1}}

\newcommand{\probpi}[2]{\mathbb{P}_{#1}\left(#2\right)}

\newcommand{\expectpi}[2]{\mathbb{E}_{#1}\left[#2\right]}

\newcommand{\condexpectpi}[3]{\mathbb{E}_{#1}\left[#2\,\middle\vert\,#3\right]}
\newcommand{\condprobpi}[3]{\mathbb{P}_{#1}\left(#2\,\middle\vert\,#3\right)}

\newcommand{\eo}[2]{\ifthenelse{\isempty{#1}}{\mathbb{Y}\left(#2\right)}{\mathbb{Y}_{#1}\left(#2\right)}}
\newcommand{\eaor}[2]{\ifthenelse{\isempty{#1}}{\overrightarrow{\mathbb{Y}}\left(#2\right)}{\overrightarrow{\mathbb{Y}}_{#1}\left(#2\right)}}
\newcommand{\eaol}[2]{\ifthenelse{\isempty{#1}}{\overleftarrow{\mathbb{Y}}\left(#2\right)}{\overleftarrow{\mathbb{Y}}_{#1}\left(#2\right)}}

\DeclareMathOperator{\efao}{EFAO}
\DeclareMathOperator{\afeo}{AFEO}
\DeclareMathOperator{\eao}{EAO}
\DeclareMathOperator{\aoe}{AOE}
\DeclareMathOperator{\supp}{supp}
\DeclareMathOperator{\variance}{Var}
\newcommand{\variancepi}[2]{\variance_{#1}\left(#2\right)}
\newcommand{\condvariancepi}[3]{\variance_{#1}\left(#2\,\middle\vert\,#3\right)}
\newcommand{\diffsupp}[2]{\supp^\Delta_{#1,#2}}
\newcommand{\diffsupppi}[1]{\supp^\Delta_{#1}}
\DeclareMathOperator{\argmax}{arg\,max}

\NewDocumentCommand{\drawbiclique}{mm}{%
  \begin{tikzpicture}[x=2.5cm, y=1cm, every node/.style={draw, circle, minimum size=0.25cm}]
    \foreach \i in {1,...,#1} {
      \node (u\i) at (0,{-\i + (#1+1)/2}) {};
    }
    \foreach \j in {1,...,#2} {
      \node (v\j) at (1,{-\j + (#2+1)/2}) {};
    }
    \foreach \i in {1,...,#1} {
      \foreach \j in {1,...,#2} {
        \draw (u\i) -- (v\j);
      }
    }
  \end{tikzpicture}
}

\theoremstyle{definition}
\newtheorem{definition}{Definition}
\newtheorem{proposition}{Proposition}
\newtheorem{corollary}{Corollary}[proposition]
\newtheorem*{remark}{Remark}